\documentclass[12pt,draftcls,onecolumn]{IEEEtran}

\usepackage[pdftex]{graphicx}
\usepackage[utf8]{inputenc}
\usepackage{psfrag}
\usepackage{amsthm}
\usepackage{amsmath}
\usepackage{amssymb}

\newtheorem{ExampleDef}{Example}[section]

\newcommand{\Example}[3]
{
  \begin{list}{}{
      \setlength{\leftmargin}{1em}
      }     
    \item                               
    \small                              
    \begin{ExampleDef} \rm              
      {\bf \hspace{-1ex}: #1}           
      #2                                
      \label{ex#3}                      
    \end{ExampleDef}
  \end{list}
  }

\begin{document}
\title{Nonlinear Contraction Tools \\ for Constrained Optimization}

\author{Jonathan Soto and Jean-Jacques E. Slotine}

\markboth{Preprint submitted to IEEE Transactions on Automatic control}%
{}

\maketitle

\begin{abstract}
 This paper describes new results linking constrained optimization
theory and nonlinear contraction analysis.  Generalizations of
Lagrange parameters are derived based on projecting system dynamics on
the tangent space of possibly time-varying constraints. The paper
formalizes the intuition that, just as convexity rather than linearity
is the key property in optimization, contraction rather than linearity
is the key dynamical property in this context.
\end{abstract}


\section{Introduction}
After presenting some initial examples that show the relations between contraction
and optimization, we derive a contraction theorem for nonlinear systems with equality
constraints. The method is applied to examples in differential geometry and biological
systems. A new physical interpretation of Lagrange parameters is provided. In
the autonomous case, we derive a new algorithm to solve minimization problems.
Next, we state a contraction theorem for nonlinear systems with inequality constraints.
Finally, we state another contraction theorem for nonlinear systems with time-varying
equality constraints. A new generalization of time varying Lagrange parameters is
given. In the autonomous case, we provide a solution for a new class of optimization
problems, minimization with time-varying constraints.

In the following we consider an n-dimensional time-varying system of the
form:
\begin{equation}\label{eq:nonlinear}
\dot {\bf x}(t) = {\bf f}({\bf x}(t), t) 
\end{equation}
where ${\bf x} \in R^n$ and $t \in R^+$ and f is $n \times 1$ nonlinear vector
function which is assumed to be real and smooth in the
sense that all required derivatives exist and are continuous.

\section{Preliminaries}
\subsection{On contraction analysis}
 Our results are based on nonlinear contraction theory \cite{oncontraction}, a
viewpoint on incremental stability whose basic result we now recall.
Historically, ideas closely related to contraction can be traced back
to \cite{Hartman} and even to \cite{Lewis} (see also \cite{Angeli}, \cite{Pavlov}, and e.g. \cite{International} for a
more exhaustive list of related references). As pointed out in \cite{oncontraction},
contraction is preserved through a large variety of systems
combinations, and in particular it represents a natural tool for the
study and design of nonlinear state observers, and by extension, of
synchronization mechanisms \cite{Cybernetics}.

The basic result of contraction analysis can be
stated as (for more details we refer to \cite{oncontraction}):

\newtheorem{theorem}{Theorem}
\begin{theorem} \label{th:contraction}
	Denote by $\frac{\partial \bf f} {\partial \bf x}$ the Jacobian
        matrix of $\bf f$ with respect to $\bf x$.  Assume that there exists a
        complex square matrix $\bf \Theta(x,t)$ such that the Hermitian
        matrix $\bf \Theta(x,t)^{*T}\bf \Theta(x,t)$ is uniformly positive
        definite, and the Hermitian part ${\bf F}_H$ of the matrix 
	\[
        \bf F = \left( \dot {\bf \Theta} + {\bf \Theta} \frac{\partial \bf f}{\partial \bf x}
        \right) {\bf \Theta^{-1}} 
	 \]
        is uniformly negative definite. Then, all system trajectories converge
        exponentially to a single trajectory, with convergence rate
        $|\sup_{\bf x,t}\lambda_{max}({\bf F}_H)|>0$. The system is said to
        be \emph{contracting}, $\bf F$ is called its \emph{generalized
        Jacobian}, and $\bf \Theta(x,t)^{*T}\bf \Theta(x,t)$ its contraction
        \emph{metric}. The contraction rate is the absolute value of the largest eigenvalue (closest to zero, although still negative) $\lambda = | \lambda_{max}(\bf{F}_H) |$. 
\end{theorem}

\subsection{Optimization theory}
Optimization means finding "best available" values of some objective function given a defined domain, including a variety of different types of objective functions and different types of domains.  We give the sufficient conditions for the point ${\bf x}^*$ to be a minimum of a minimization problem. There are different classes of problems which are stated below. (for more details we refer to \cite{Bertsekas})

In unconstrained optimization we search for the minimum over the whole range of variables. The generic form is:
\begin{equation}\label{eq:unconstrained}
min\ U({\bf x})
\end{equation}
The second-order sufficient conditions in the unconstrained case are given by theorem \ref{th:unconstrained}.

\begin{theorem} \label{th:unconstrained}
Let $U \in C^2(R^n,R)$ be a function defined on a region in which the point ${\bf x}^*$ is an interior point. Suppose in addition that 
$$\nabla U({\bf x}^*) = 0 \ \ and \ \nabla^2 U({\bf x}^*)\ is\ positive\ definite$$
$x^*$ is a solution of problem \ref{eq:unconstrained}.
\end{theorem} 

In constrained optimization, the Karush-Kuhn-Tucker conditions (also known as the Kuhn-Tucker or KKT conditions) are sufficient for a solution in nonlinear programming to be optimal, provided that some regularity conditions are satisfied.
The generic form is with m constraints is:
\begin{equation}\label{eq:constrained}
min\ U({\bf x}) \ \  subject\ to\ (s.t)\ \ {\bf h}({\bf x})={\bf 0}
\end{equation}
The second-order sufficient conditions in the constrained case are given by theorem \ref{th:constrained}.
\begin{theorem} \label{th:constrained}
 {Let $U({\bf x}) \in C^2(R^n,R)$ and ${\bf h}({\bf x}) \in C^2(R^n,R^m)$ be smooth functions and ${\bf x}^*$ is a regular point (the columns of $\nabla {\bf h}({\bf x^*}) $ are linearly independent).
Suppose in addition that 
$$\nabla L({\bf x}^*) = 0  \ and \ {\bf y}'\nabla^2 L({\bf x}^*){\bf y}>0 \ $$
where ${\bf y}\nabla {\bf h}({\bf x}^*)=0$, $\lambda$ are the Lagrange parameters and $L $ is the lagrangian function defined as $L=U+{\bf \lambda} {\bf h}$.\\
${\bf x}^*$ is a solution of problem \ref{eq:constrained}.
}
\end{theorem} 
\subsection{Examples}
We present some cases - unconstrained optimization, duality theory - in which both theories are linked.
The following example in \cite{modular} is the starting point of this research.
\Example{}{\label{ex:link}
Consider a gradient autonomous system $\dot{\bf x}=-\nabla U({\bf x})$, contracting in an identity metric, $\Theta = I$. As it is autonomous and contracting in a time independent metric, it has a unique equilibrium point because
$$
\frac{d}{d{t}}( \nabla U) = {\bf F} ( \nabla U)
$$
which implies exponential convergence of $\dot{\bf x}$ to zero. ${\bf x}$ converges to a constant, ${\bf x}^*$.
This point has zero speed, $\nabla U({\bf x}^*)=0$. 
The condition of contraction at this equilibrium point is the positive definiteness of
$\nabla^2 U({\bf x}^*)$.
These are the two sufficient conditions (theorem \ref{th:unconstrained}) to prove that ${\bf x}^*$ is a solution to problem \ref{eq:unconstrained} 
}{1}

Contraction theory is also related to differential geometry through example \ref{ex:dgeometry}. \cite{hamiltonian}
\Example{}{\label{ex:dgeometry}
The condition of contraction of the system \ref{eq:nonlinear} for a positive definite metric ${{\bf g}={\bf \Theta}'{\bf \Theta}}$ that verifies a parallel propagation is that the hermitian part of
$$ {\bf F} = {\bf \Theta} D{\bf f}({\bf x},t) {\bf \Theta}^{-1} = {\bf \Theta} \left (\frac{\partial{\bf f}({\bf x},t)}{\partial{\bf x}} +\sum_i{\bf \Gamma}^k_{ij} {\bf f}^i({\bf x},t)\right ){\bf \Theta}^{-1} $$
has a uniformly negative maximum eigenvalue.
$D$ is the covariant derivative and the Christoffel term is defined as $\sum_k \Gamma^k_{ij}g_{kl}=\frac{1}{2}(\frac{\partial g_{il}}{\partial q^j}+\frac{\partial g_{jl}}{\partial q^i}-\frac{\partial g_{ij}}{\partial q^l})$.

The idea is to use a parallel propagation of the tensor $ {\bf \Theta}$, \cite{hamiltonian}. 
$$
\dot{ {\bf \Theta}} ={\bf \Theta} \sum_i {\bf \Gamma}^k_{ij} \dot{{\bf x}^i}  
$$
}{2}
The geometric interpretation of the covariant derivative is the projection of the directional derivative \cite{Kuhnel} on that submanifold, thus the tangential part of the directional derivative. It can be written as:
$$
D_X Y = (d_X Y)^{tang}=d_X Y-<d_X Y,\nu>\nu
$$
where $d_X Y$ is the directional derivative along the X direction and  $\nu$ is the normal vector to the submanifold. \\
When writing $\delta \dot z = \Theta D {\bf f} \Theta^{-1}\delta z$, it means that the virtual speed is constrained to belong to 'some' manifold. This idea of projection is fundamental to prove contraction for constrained dynamical systems. 

With example \ref{ex:link}, we show that contraction and unconstrained optimization are linked through gradient systems. Another important domain in optimization is duality. It is very useful to use the dual formulation instead of the primal for a variety of reasons. Sometimes the dual problem has a closed form solution or the algorithm to find the minimum is much faster. In example \ref{ex:dual}, we show that contraction and duality are related using gradient systems. 

The Legendre transformation is defined as follows for $y\in R^n$:
\begin{equation}\label{eq:dual}
U^*({\bf y})=sup_{{\bf x}\in R^n}({\bf x}'{\bf y}-U({\bf x}))
\end{equation}
The conjugate function $U^*$ is a convex function since it is the pointwise supremum of linear functions. 

\Example{}{\label{ex:dual}
The system $ \dot{{\bf x}} = - \nabla{U({\bf x})}$ is contracting for the identity metric and also smooth (existence of derivatives) if and only if
the system $ \dot{{\bf y}} = - \nabla{U^*({\bf y})}$ is contracting for the identity metric and also smooth.\\
To prove the if part we use the fact that strict convexity and smoothness are dual 
properties, \cite{dual}. \\
For the only if part, we apply the first part of the proof to the function $U^*$. We use the fact that as $U^*$ convex, close (the domain of U is closed) and proper (it never takes the value $-\infty$ and the set $dom_g = [x | g(x) < \infty]$ is nonempty), $U^{**} =U$
}{3}
In \cite{oncontraction} combination properties of contracting systems are of great importance. In example \ref{ex:combination} we derive a new result that cannot be achieved with classical minimization theory.

\Example{}{\label{ex:combination}
If there are two unconstrained minimizations which have respectively their minimum at $x_1^*$ and $x_2^*$, the sum of both cost functions may not have a minimum. Using only classical minimizations theorems we cannot conclude anything when summing both systems. 

If the gradient systems issued from the two precedent minimization problems are contracting, the sum of the gradient systems, is still contracting. Therefore the sum of the two cost functions has a minimum
}{4}

In the following all the points are regular. This means that $\nabla {\bf h}({\bf x}) \nabla {\bf h}'({\bf x})$ has full rank and hence is invertible. That is the {\it only} hypothesis needed in the following analysis.

\section{Contraction theory with equality constraints}
\subsection{Starting on the constraints: contraction theory}
We define the sets $S = [ {\bf x}\ / \ {\bf h}({\bf x})=0]$ and $M({\bf x})=[ y\ / \ y\nabla {\bf h}({\bf x})=0]$. S is the space of the constraints. M is the tangent subspace of {\bf h} at point ${\bf x}$, it is a k-submanifold. (a k submanifold is defined as $ \nabla {\bf h}({\bf x})$ having full rank using the implicit function theorem).

We define the operator \cite{luenberger}
$$P({\bf x})=1-\nabla {\bf h}'({\bf x})[{\nabla {\bf h}({\bf x}) \nabla {\bf h}({\bf x})'}]^{-1}\nabla {\bf h}({\bf x})$$
It is a symmetric orthogonal projection operator onto $M({\bf x})$ because it verifies $P({\bf x})P({\bf x})=P({\bf x})$ and $P({\bf x})y \in M({\bf x})$. 
\subsubsection{Contraction theorem }
\begin{theorem}\label{th:equality}
The condition of contraction of the constrained dynamic system $$\dot {\bf x} = {\bf f}({\bf x},t)  \  s.t. \ {\bf h}({\bf x})=0$$ is that there exists a metric, $\Theta$, uniformly positive definite and that the hermitian part of :
$$
{\bf F} = \left (\dot {\bf \Theta}({\bf x},t)  +  {\bf \Theta}({\bf x},t)P({\bf x})\left (\frac{\partial{\bf f}}{\partial{\bf x}}+\lambda({\bf x},t)\frac{\partial^2{\bf h}}{\partial{\bf x}^2}\right ) \right ) {\bf \Theta}({\bf x},t)^{-1} \ \ on \ S({\bf x})
$$
has a maximum eigenvalue uniformly negative. $P({\bf x})$ is the projection operator onto the tangent space $M({\bf x})$ and $\lambda({\bf x},t)=-{{\bf f}({\bf x},t)\nabla {\bf h}'}[{\nabla {\bf h}\nabla {\bf h}'}]^{-1}$. The initial condition must verify the constraint.
\end{theorem}

The proof has three parts. We project the system by adding a term related to the Lagrange parameters. We compute the condition of contraction of this new system.Finally, we prove contraction behavior. 

The initial problem is 
$$\dot {\bf x} = {\bf f}({\bf x},t)  \  s.t.  \ {\bf h}({\bf x})=0$$
Physically ${\bf f}$ is the velocity of the system going away from the constraints. To verify the constraints, the systems's velocity has to be tangent to the constraints at {\bf x}. We project the system onto M({\bf x}) using the projector P({\bf x}).
$$\dot {\bf x} =f({\bf x},t)+\lambda({\bf x},t)\nabla {\bf h}({\bf x})$$
We note $\lambda({\bf x},t)=-{f({\bf x},t)\nabla {\bf h}'({\bf x})}[{\nabla {\bf h}({\bf x}) \nabla {\bf h}({\bf x})'}]^{-1}$. These are not exactly the Lagrange parameters. But if ${\bf f}= \nabla U({\bf x})$, we recognize the Lagrange parameters. This is a first generalization of the Lagrange parameters. The new dynamical system is $\dot {\bf x} = \nabla U({\bf x})+\lambda({\bf x})\nabla {\bf h}({\bf x})=\nabla { L({\bf x})} $ where L is the usual Lagrangian function. In the following the projected system will be known as Lagrangian dynamics.
\begin{figure} [ht]
\centering
\includegraphics[scale=0.75]{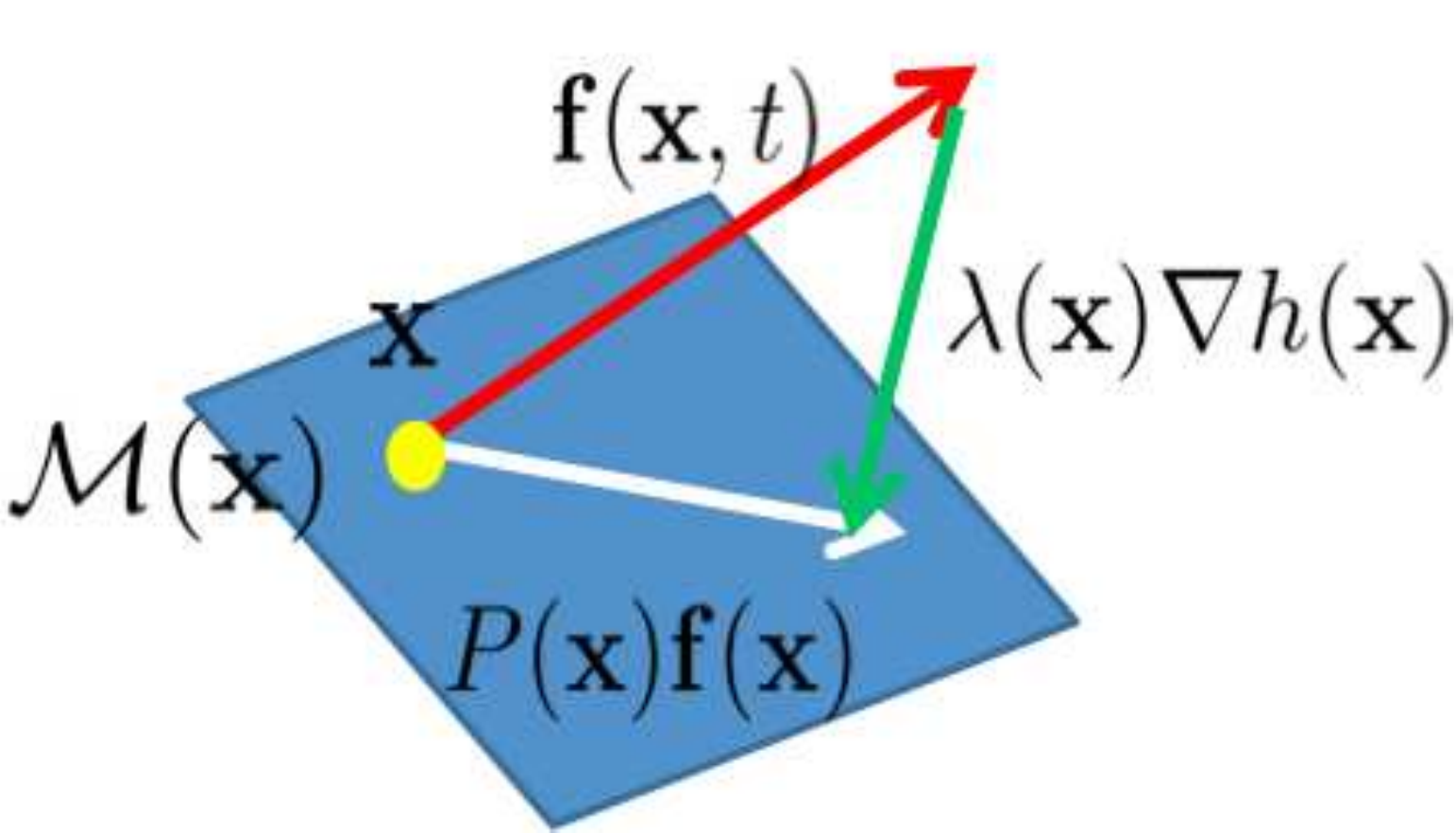}
\caption{Projected speed of the dynamic system}
\label{fig:pictures/pic.pdf}
\end{figure}
This result gives a new insight about Lagrange parameters. The term $\lambda({\bf x},t)\nabla {\bf h}({\bf x})$ is a reaction force allowing the system to stay on the constraints. A Lagrange parameter is the scalar value that gives the magnitude of the force along the orthogonal direction to a particular constraint in order to have a tangential speed. Lagrange parameters as reaction forces has already been investigated using the Lagrange equation \cite{Gignoux} \cite{EricFeron}.
Anytime, the constraints are verified because the system starts on the constraints and the velocity is tangential to the constraints. If initially the constraints were not verified, this property would not be true.

We consider a virtual displacement $\delta {\bf x}$ and a positive definite metric, ${\bf \Theta}({\bf x},t)$. 
The function $\delta {\bf x} '{\bf \Theta}({\bf x},t)'{\bf \Theta}({\bf x},t)\delta {\bf x} $ is the distance associated to the first fundamental form of the k-submanifold $M({\bf x})$. We define $\delta {\bf z} = {\bf \Theta}({\bf x},t)\delta {\bf x}$.

We compute $\frac{d}{dt}(\delta {\bf {\bf z}}' \delta {\bf {\bf z}})  = 2\delta {\bf {\bf z}}' \delta \dot {\bf {\bf z}}$. We calculate $\delta \dot {\bf z} = \dot {\bf \Theta}({\bf x},t) \delta {\bf x} + {\bf \Theta}({\bf x},t) \delta \dot{{\bf x} }$ where $ \delta \dot{{\bf x} } =\left ( \frac{\partial f}{\partial{\bf x}}+\lambda({\bf x},t)\frac{\partial^2{\bf h}}{\partial{\bf x}^2}+ \frac{\partial\lambda({\bf x},t)}{\partial{\bf x}}\frac{\partial{\bf {\bf h}}}{\partial{\bf x}}\right ) \delta {\bf x}$

We compute a long calculation $\frac{\partial\lambda({\bf x},t)}{\partial{\bf x}}$. This is done in \cite{luenberger} and \cite{MyThesis}.
\begin{equation}\label{eq:2parts}
\begin{array}{ll}
 \delta \dot{{\bf x} } =  P({\bf x})\left (\frac{df}{d{\bf x}}+\lambda({\bf x},t)\frac{\partial^2{\bf h}}{\partial{\bf x}^2}\right ) \delta {\bf x} -{\nabla {\bf h}({\bf x})'}[{\nabla {\bf h}({\bf x}) \nabla {\bf h}({\bf x})'} ]^{-1}\delta {\bf x}\frac{\partial^2{\bf h}}{\partial{\bf x}^2}\dot {\bf x}
\end{array} 
\end{equation}
The first term belongs to $M({\bf x})$, the second term to $M({\bf x})^{\perp}$. The equation can be rewritten :
$$ \delta \dot{{\bf x} } = \delta \dot {\bf x}^{\parallel}+ \delta \dot {\bf x}^{\perp} $$ In order to have contraction behavior the first term of $\delta \dot{{\bf x} }^{\parallel}$ must be uniformly bounded. 
Finally 
$$
\frac{d}{dt}(\delta {\bf z}' \delta {\bf z})  = 2\delta {\bf z}' \left (\dot {\bf \Theta}({\bf x},t) + {\bf \Theta}({\bf x},t)  P({\bf x}) \left (\frac{\partial f}{\partial{\bf x}}+\lambda({\bf x},t)\frac{\partial^2{\bf h}}{\partial{\bf x}^2}\right )\right) {\bf \Theta}({\bf x},t)^{-1}\delta {\bf z} 
$$
\Example{}{
The major goal of control theory is to find the control input that will make the system behave in a defined way. In mathematical terms, find $\hat {\bf u}$ such that 
$$
\dot{\bf x} = {\bf f}({\bf x},t)+{\bf u}({\bf x},t) 
$$
has a specific behavior. In the case of projected contraction theory identifying $\hat {\bf u}({\bf x},t) ={\bf \lambda }\nabla {\bf h}$, we find a control that makes the system evolve on the constraint. 
}{5}
The following example shows how projected contraction can be applied to general dynamic systems (not only to gradient systems, i.e. minimization problems).\\
\Example{}{
The system \ref{eq:xdef} arises from the study of a mathematical model for the spread of an
infectious disease in a population with a fixed total size. The variables s, e, i, and r represent
fractions of the population that are susceptible, exposed (in the latent
period), infectious, and recovered, respectively. All parameters are assumed
to be nonnegative, we assume also $\epsilon > 0$ and $\gamma>0$. 
\begin{equation}
\begin{array}{ll}
\dot s = b-bs-\lambda is+\alpha is +\delta r\\
\dot e =\lambda is -(\epsilon+b)e+\alpha ie \\
\dot i = \epsilon e - (\gamma+\alpha+b)i+\alpha i^2\\
\dot r = \gamma i -(b+\delta)r + \alpha ir\\
\end{array} 
\label{eq:xdef}
\end{equation}
The biological feasible region is the following invariant simplex $s+e+i+r=1$, the sum of the different populations is the total population.
We project the dynamic system. The value of $\lambda=-f \nabla h'[\nabla h\nabla h']^{-1}=-\frac{1}{4}(b-bs+\alpha is  -be+\alpha ie - (\alpha+b)i+\alpha i^2 -br + \alpha ir)$.
The jacobian of the system is:
$$\frac{df}{dx}=
 \left( \begin{array}{cccc}
 -b-\lambda i+\alpha i & 0 & (-\lambda+\alpha)s & \delta \\
       \lambda i & -(\epsilon+b)+\alpha i & \lambda s+\alpha i & 0\\
       0 & \epsilon & -(\gamma+\alpha+b)+\alpha i & 0\\
       0 & 0 & \alpha e & -(b+\delta)+\alpha i\\
 \end{array} \right)
$$
The projection matrix is :
 $$
 P=1-\nabla h'({\bf x})[{\nabla h({\bf x}) \nabla h({\bf x})'}]^{-1}\nabla h({\bf x})=\frac{1}{4}
 \left( \begin{array}{cccc}
 3& -1 & -1& -1 \\
       -1& 3 &-1& -1\\
       -1 & -1 & 3 & -1\\
       -1 & -1 & -1 & 3\\
 \end{array} \right)
 $$
On figure \ref{fig:epidemic} 
, the system converges to ${\bf x}^*= [0.050,0.542,0.193,0.213 ]$. This equilibrium point corresponds to an endemic population.
The condition of contraction has to be calculated for a metric, $\Theta( {\bf x})$. In this case the metric is the complex matrix that diagonalizes the jacobian times the projection matrix.
 $$
F = \Theta( {\bf x}) P\frac{df}{dx}\Theta( {\bf x})^{-1} 
$$
We do not have $\lambda \nabla^2 {\bf h}$ in the contraction condition because the constraint is linear. There is just one eigenvalue that is equal to zero. We get rid of it by using a three times four projection matrix. The vectors of this matrix are a basis for M. Using the interlacing theorem, the eigenvalues of the projected matrix are all negative, the system is contracting.
\begin{figure} [ht]
\centering
\includegraphics[scale=0.75]{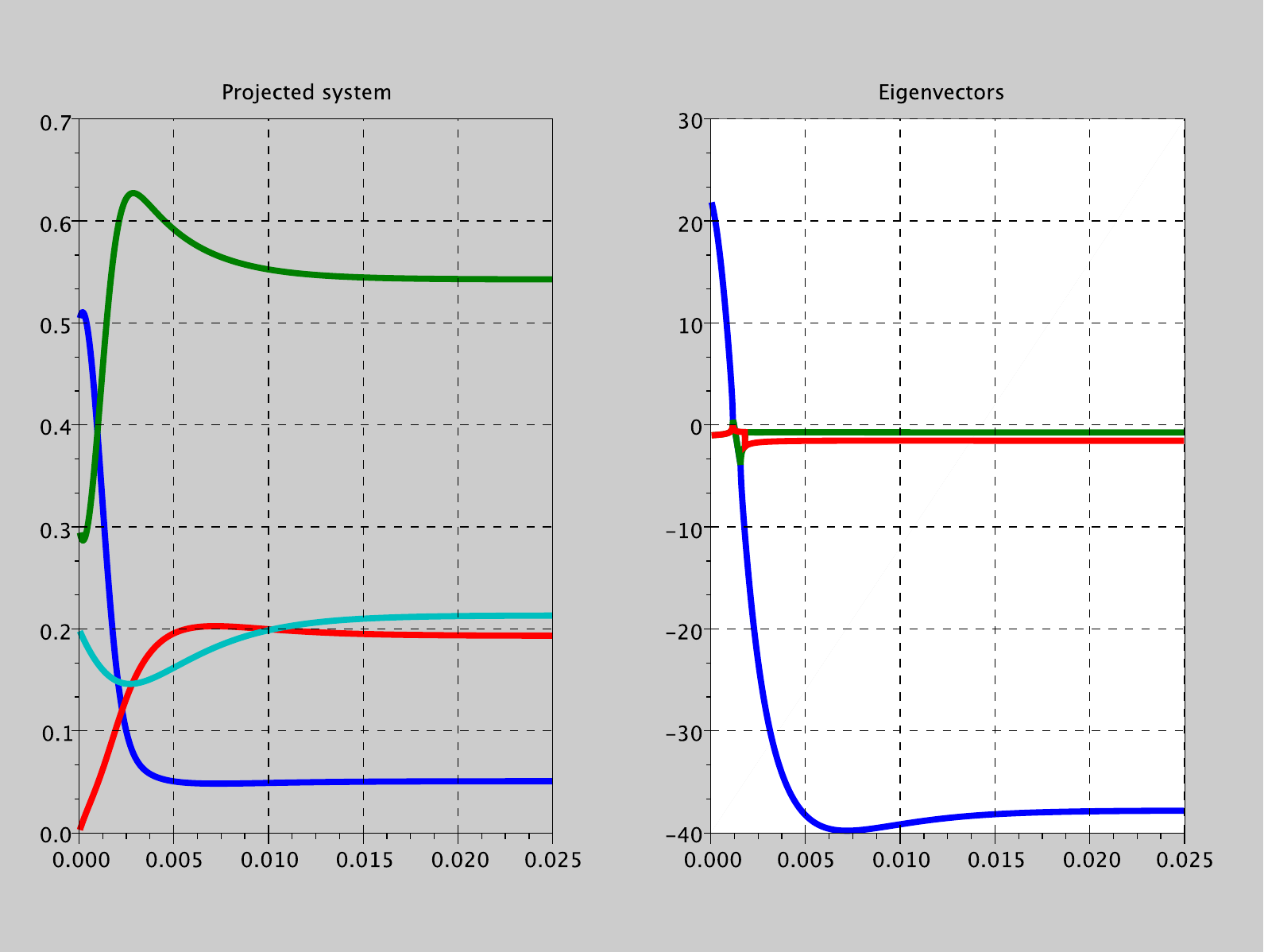}
\caption{Projected speed of the dynamic system}
\label{fig:epidemic}
\end{figure}
}{6}
\subsubsection{An algorithm for solving minimization problems}
As shown in example \ref{ex:link} an autonomous contracting dynamic system has a unique equilibrium point ${\bf x}^*$. 
\begin{equation}\label{eq:gradientconstrained}
\dot {\bf x} =- \nabla U({\bf x}) \ s.t.\  {\bf h}({\bf x})=0
\end{equation}
The projected system, which we call Lagrangian dynamic, is $\dot {\bf x} =- \nabla L({\bf x})$. It verifies $ \nabla L({\bf x}^*)=0 $.
The dynamic system also satisfies the constraint
$ {\bf h}({\bf x}^*)=0 $
because we start on the constraint and the speed of the system is always tangential to the constraints. These two conditions ensure us that at anytime ${\bf h}({\bf x})=0$.

The system being autonomous, the condition of contraction is, \ref{th:equality} $$ \ddot{\bf x}=-P({\bf x})\nabla^2 L({\bf x})\dot {\bf x} $$
As  $ \dot {\bf x} \in M({\bf x})$, we have : $ \dot {\bf x}=P({\bf x})\dot {\bf x}$. Thus:
$\ddot{\bf x}=-P({\bf x})\nabla^2 L({\bf x})P({\bf x})\dot {\bf x} $. The condition of contraction is 
$$
{{\bf x}^*} ' P'({\bf x}^*) \left (\frac{d^2U}{d{\bf x}^2}+\lambda({\bf x}^*)\frac{\partial^2{\bf h}}{\partial{\bf x}^2}\right ) P({\bf x}^*){\bf x}^*={\bf y}' \nabla^2 L({\bf x}^*) {\bf y}  > \eta>0
$$
with ${\bf y}={\bf x} P({\bf x}^*) \in M({\bf x}^*)$ by definition of $P({\bf x}^*)$.

These are the three conditions for the existence of a minimum to problem \ref{eq:constrained}.
This is a new relationship between minimization and dynamical systems using contraction theory. This has been investigated \cite{Absil}.

This idea of projection creates a new dynamical system. It is a new way of solving minimization problems. We give some examples. The first one is done in great detail.

\Example{}{
The minimization of the length of a square in a circle is
$$
min  \ x+y \ s.t. \ x^2+y^2=1
$$
The gradient dynamical associated is
$
\dot x = - \nabla f =\left( \begin{array}{ccccc} -1 \\ -1 \end{array} \right)
\ s.t. \ x^2+y^2=1
$. 
After projection
$$
\dot x =
\left( \begin{array}{ccccc} -1 +x^2+xy \\  -1+y^2+xy \end{array} \right)
$$
where $\lambda = -\frac{\nabla f \nabla h'}{\nabla h \nabla h'} = -\frac{x+y}{2}$. 
The eigenvalue of $F$ is $x+y$.
The system is contracting in $x+y \le 0$. 
As the system approaches the solution, $\lambda \rightarrow \lambda^*$. Figure \ref{fig:mincircle2}
shows that the dynamic system always stays on the constraint while it goes to the minimum.

\begin{figure} [ht!]
\centering
\includegraphics[scale=0.75]{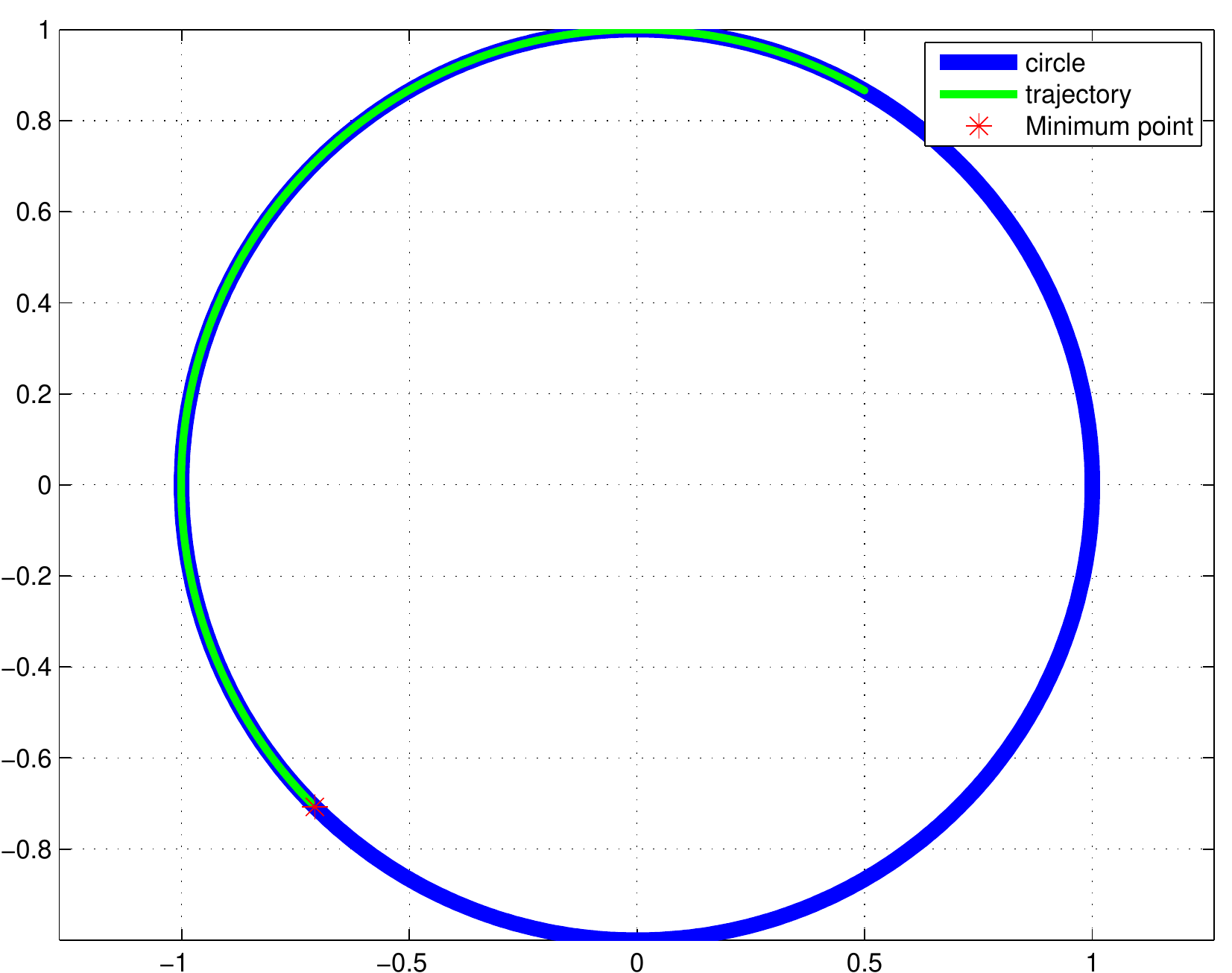}
\caption{Minimization of the length of a square in a circle in 2D}
\label{fig:mincircle2}
\end{figure}

}{7}

\Example{}{
Consider the minimum distance from a point to a sphere problem 
$$
min\ \frac{1}{2}(x^2+y^2+(z-2)^2) \ \ \ s.t. \ \ h(x)=x^2+y^2+z^2-1=0
$$
The evident solution of this problem is the projection of the point $(0,0,2)$ on the sphere: $(0,0,1)$. The projected system is:
$$
\dot x = \nabla U +\lambda \nabla h=
\left( \begin{array}{ccccc} 
x+\lambda x\\
y+\lambda y \\
z-2+\lambda z \\
 \end{array} \right)
$$
with $\lambda=2z-1$. 

On figure 
\ref{fig:sphere1} 
, the system evolves on the constraint. it converges towards the expected solution. 

\begin{figure} [ht]
\centering
\includegraphics[scale=0.75]{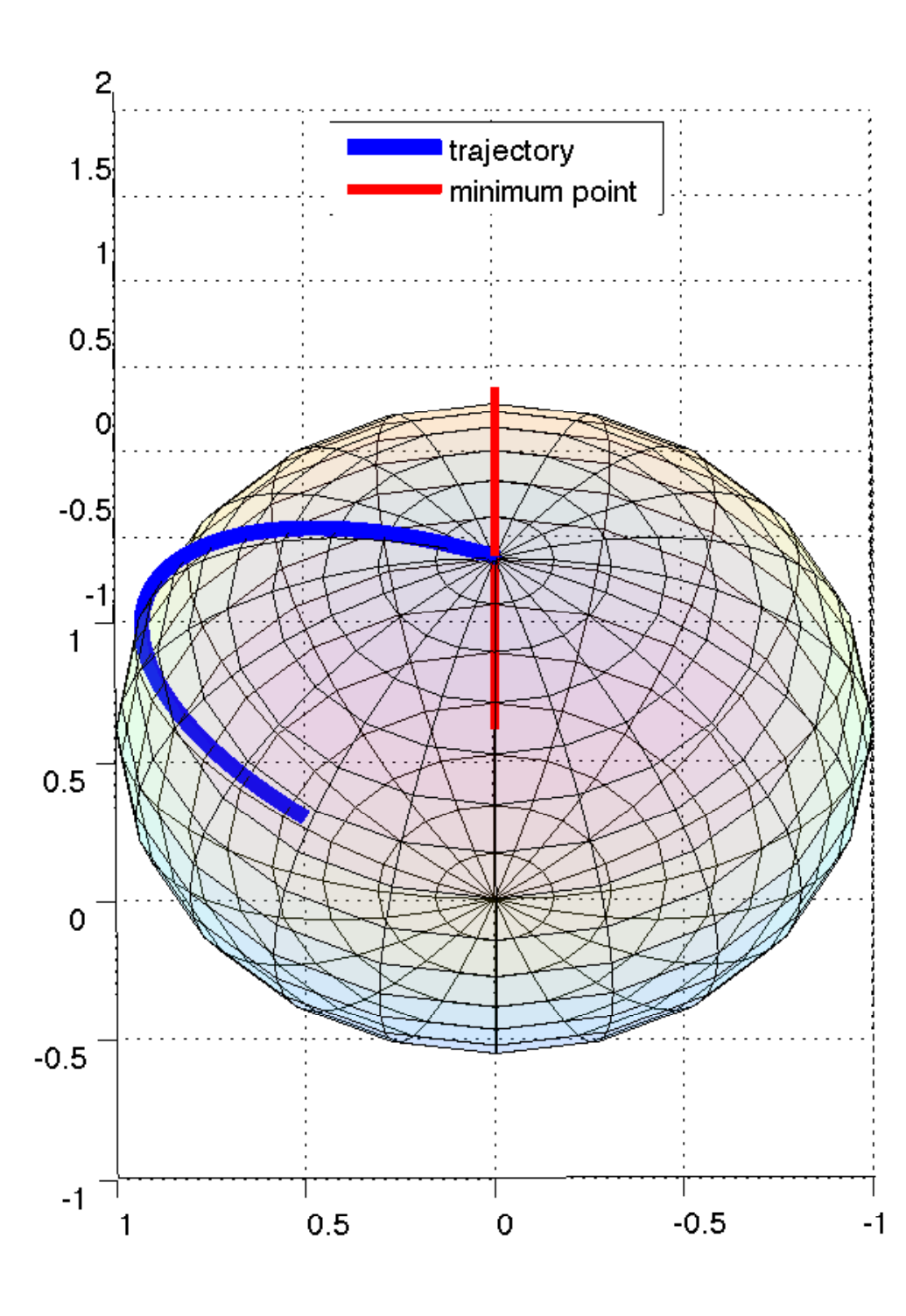}
\caption{Minimum distance from a point to a sphere}
\label{fig:sphere1}
\end{figure}

}{8}

\Example{}{The minimum distance from a point to an ellipse
 is chosen to show that the lack of symmetry does not prevent convergence of the algorithm. We search the distance from a point to the ellipse.
$$
min\ \frac{1}{2}||{\bf x}-{\bf x_1}||^2 \ \ \ s.t. \ \ h({\bf x})=\frac{x^2}{a^2}+\frac{y^2}{b^2}+\frac{z^2}{c^2}-1=0
$$
Consider $a=15;b=5;c=3;$ and the absolute minimum being $1,4,2$. On figure \ref{fig:ellipsedistance} 
the solution corresponds to the projection of the absolute minimum on the ellipsoid. (The red dotted line just indicates where the projection point is).
\begin{figure} [ht]
\centering
\includegraphics[scale=0.75]{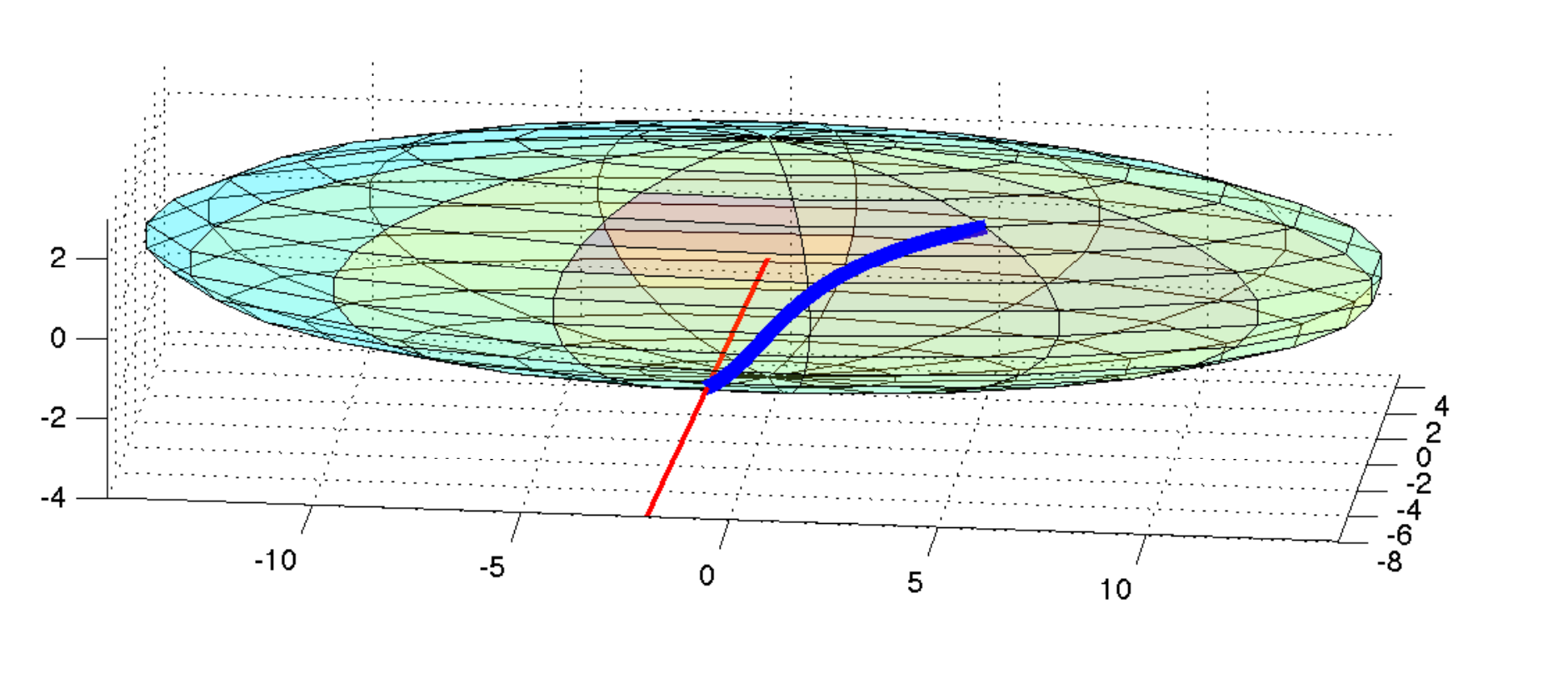}
\caption{Dynamical system on the ellipsoid, starting point is  $6.7,-2.5,2.2$ }
\label{fig:ellipsedistance}
\end{figure}
As usual the convergence's speed is very fast compared to matlab.
}{9}

\Example{}{
The minimization problem with two constraints
$$
min\ \frac{1}{2}||{\bf x}-{\bf x_1}||^2  \ \ \ s.t. \ \ h_1({\bf x})=||{\bf x}||^2-1=0 \  \ h_2({\bf x})=x+y+z=0
$$
The absolute minimum is $1,4,2$. The starting point has to verify both constraints are verified : $[0,\frac{1}{\sqrt{2}},-\frac{1}{\sqrt{2}}]$. The dynamical system follows both constraints. Each Lagrange parameter acts as a force making the dynamical system to stay on the corresponding constraint. The final point is the double projection of the absolute minimum, figure \ref{fig:twoconstraints} 
\begin{figure} [ht]
\centering
\includegraphics[scale=0.75]{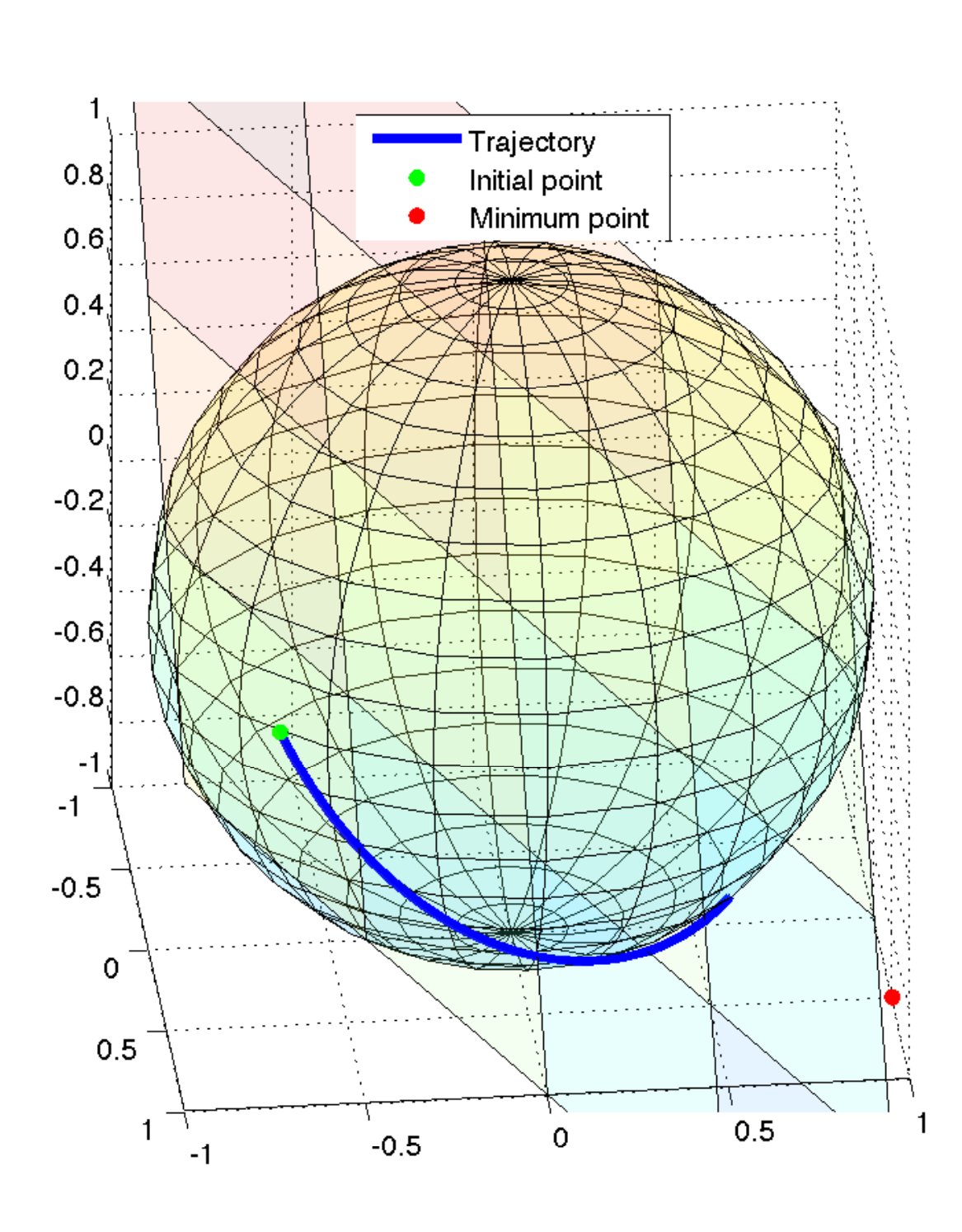}
\caption{Example of minimization with two constraints}
\label{fig:twoconstraints}
\end{figure}
}{10}

\Example{}{Consider the minimization on a torus 
$$
min\ x \ \ s.t. \ h({\bf x})=(R-\sqrt{x^2+y^2})^2+z^2-r^2=0
$$
The minimum on the torus is on one of the sides. On figure \ref{fig:torus} 
, the system evolves on the constraint even if the shape of the surface is  complicated. R corresponds to the outer radius and r corresponds to the inner radius.
\begin{figure} [ht]
\centering
\includegraphics[scale=0.75]{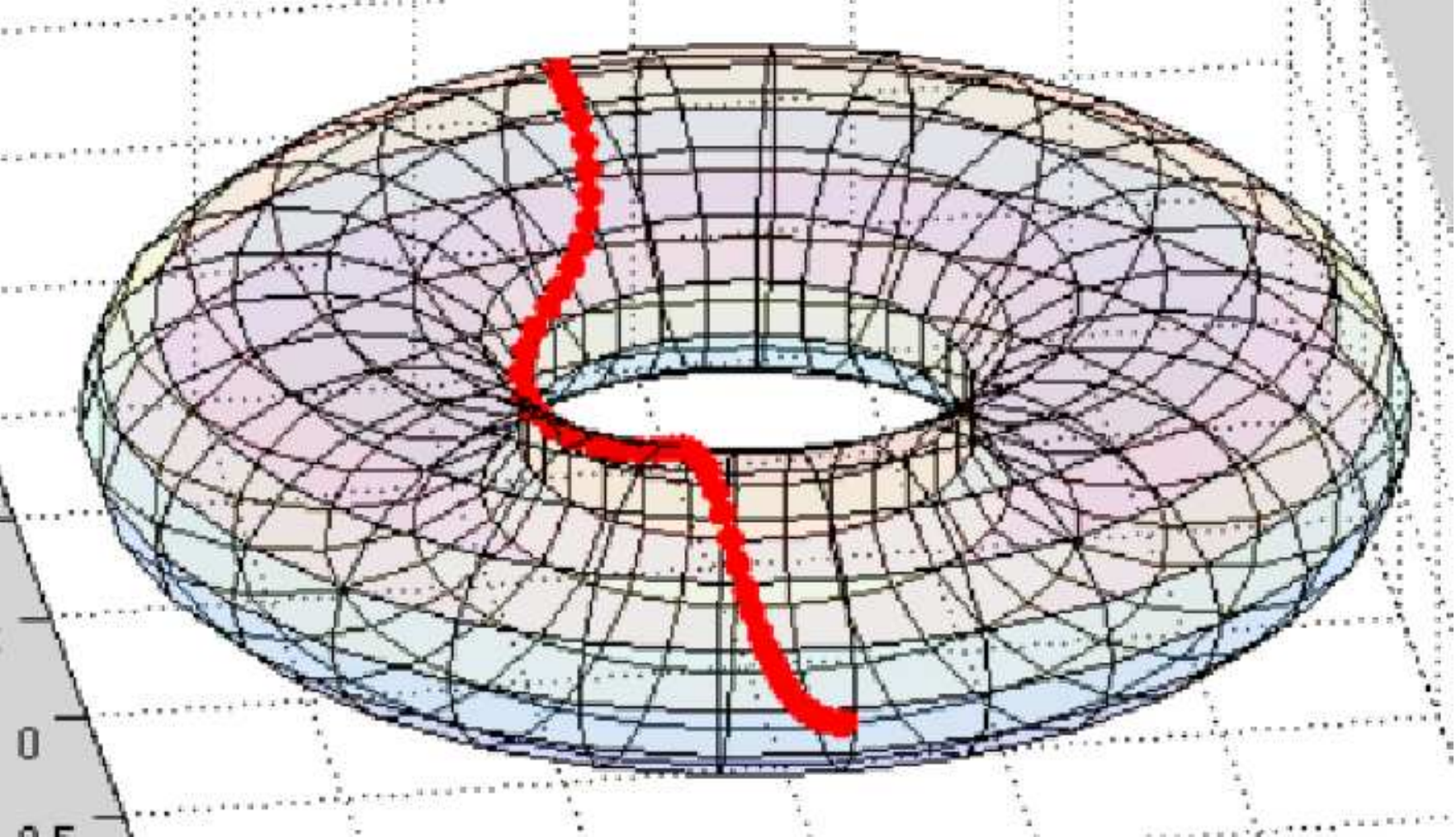}
\caption{Minimization on a torus}
\label{fig:torus}
\end{figure}
}{11}

\subsection{Starting outside the constraints: sliding behavior}\label{ch2: outsidetheconstraints}
\subsubsection{The problem and intuition of the solution}
\begin{figure} [ht]
\centering
\includegraphics[scale=0.75]{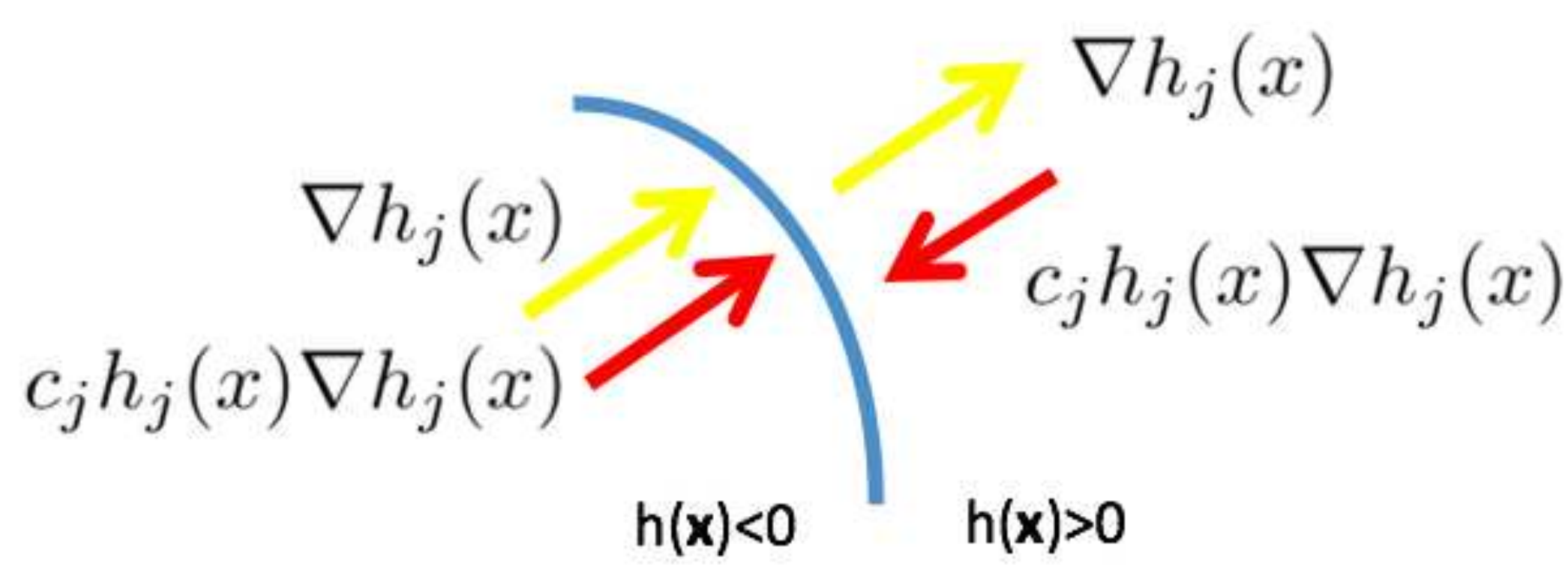}
\caption{Sliding dynamic}
\label{fig:sliding}
\end{figure}

The biggest limitation of this problem is to have to start on the constraints. One solution is to add the following dynamic.
$$
-\sum_i c_i h_i({\bf x})\nabla h_i({\bf x})
$$
$c_i$ is a constant that gives the speed of approach to the constraint i. $\nabla h_i({\bf x})$ gives the direction that is normal to the surface locally and $-h_i({\bf x})$ gives a sense to this direction, figure \ref{fig:sliding} 
. A priori, this property is local because far away from the surface the term in $\nabla h_i$ is not necessarily normal to the surface. 
Consider the modified dynamic system:
$$
\dot {\bf x} = f({\bf x},t)+\lambda \nabla h({\bf x}) - \sum_j c_j h_j({\bf x})\nabla h_j({\bf x}) 
$$ 
In an autonomous gradient system, 
$$
\dot {\bf x} = -\nabla (U({\bf x})+\lambda  h({\bf x}) + \frac{1}{2}\sum_j c_j h^2_j({\bf x}))=-\nabla L_c
$$
This term corresponds to the term that is added in optimization to construct the augmented lagrangian, $L_c$. It can be understood as adding a cost when the constraints are not verified. 

\subsubsection{Sliding surface behavior}
In this section we prove that converging to the surface when starting away from is a global property. We use sliding control techniques. We define ${\bf s}=\frac{1}{2}(\sum_i c_i h^2_i({\bf x}))$ where $c_i>0$. s is by definition non negative. Its derivative is
$$
\frac{d{\bf s}}{dt}=\sum_i c_i h_i({\bf x}) \nabla h_i({\bf x}) \dot {\bf x} 
$$
Because $f({\bf x},t)P({\bf x}) \in M({\bf x})$, $\nabla h_i({\bf x})f({\bf x},t)P({\bf x}) = 0$. The derivative is equal to :
$$
\frac{d{\bf s}}{dt}=-\sum_i c_i h_i({\bf x}) \nabla h_i({\bf x})( \sum_j c_j h_j({\bf x})\nabla h_j({\bf x}))=-(\sum c_i h_i \nabla h_i)^2 \le 0
$$
If $\ddot {\bf s}$ is bounded, 
we apply Barbalat's lemma \cite{slotinebook}. $\sum c_i h_i \nabla h_i$ converges to zero as time goes to infinity. Using the initial hypothesis,  $\nabla h_i $ are linearly independent, $c_i h_i$ goes to zero. The constraints are verified. The technique used here is a very common technique in nonlinear control theory. ${\bf h}^2$ can be seen as the distance to the surface, which is the same as the sliding variable ${\bf s}$ used in nonlinear control. 

With one constraint we get exponential convergence towards the surface. 
$$
\frac{d}{dt}\frac{h^2}{2}=h({\bf x}) \nabla h({\bf x}) \dot {\bf x} = - c h^2({\bf x}) (\nabla h({\bf x}))^2
$$
There are two dynamics in this augmented system. These two dynamics can be controlled using the ${\bf c}$ parameter. The first dynamic is the so called, Lagrangian dynamics. This dynamic leads to the minimum once on the surface. The second dynamic is the so called sliding dynamics. It makes the system converge to the constraints. If the value of ${\bf c}$ is very big, we give more importance to the sliding term. If ${\bf c}$ is small, it will take longer to converge.
\subsubsection{Examples}
\Example{}{
Consider the optimization problem
$$
min  \ x+y \ s.t. \ x^2+y^2=1
$$
The projected system is:
$$
\dot x =
\left( \begin{array}{ccccc} -1 +\frac{x^2+xy}{x^2+y^2}-c(x^2+y^2-1)x\\  -1+\frac{y^2+xy}{x^2+y^2}-c(x^2+y^2-1)y \end{array} \right)
$$
If c is small, (figure \ref{fig:mincirclesliding} ) initially the system follows the Lagrangian dynamic, (it follows a circle that is situated at the starting position). As the system goes away from the surface, the sliding dynamic becomes more important.
\begin{figure} [ht]
\centering
\includegraphics[scale=0.75]{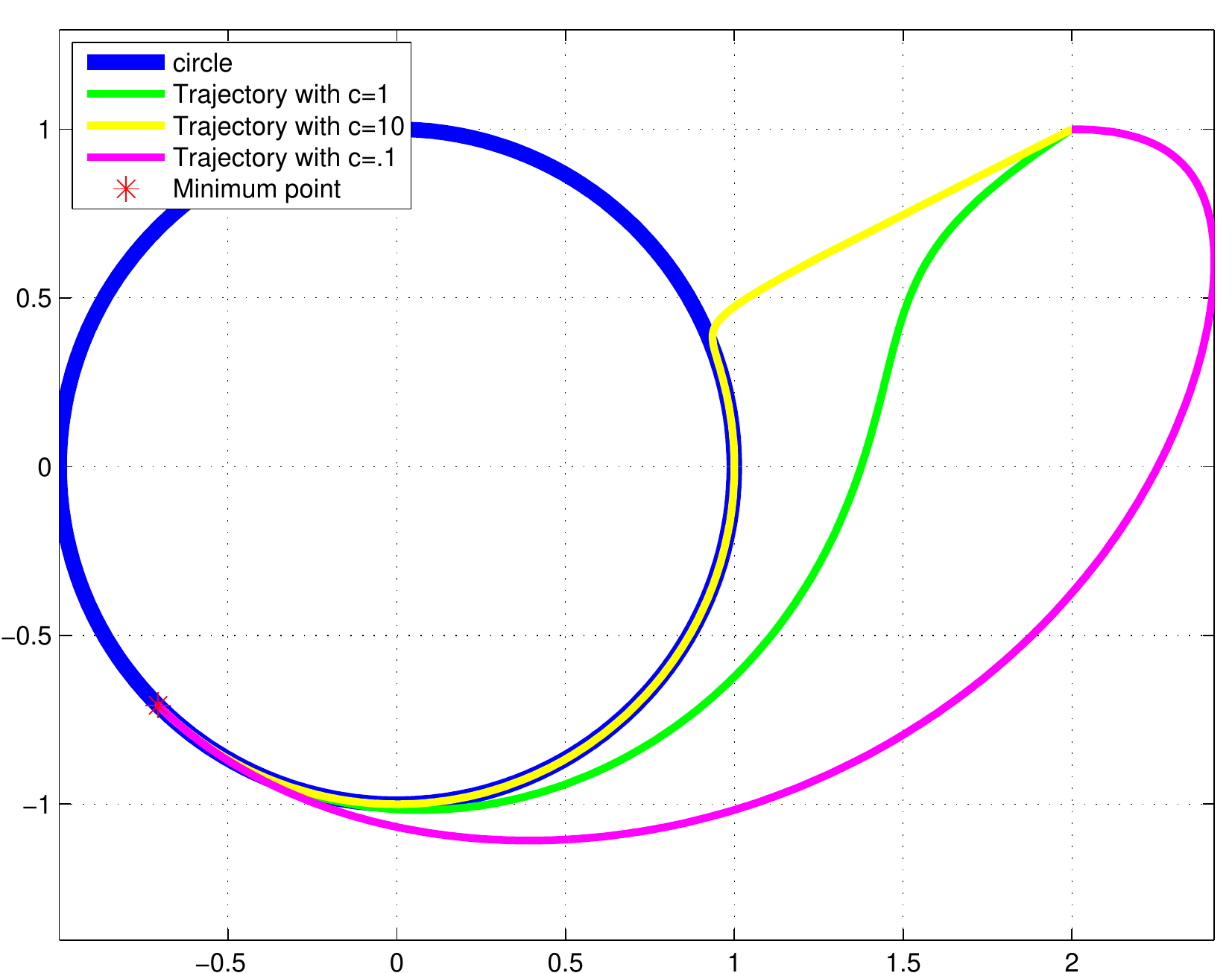}
\caption{Minimization of the length of a square in a circle}
\label{fig:mincirclesliding}
\end{figure}
If c is big, the system converges very quickly to the surface. Once on the surface the Lagrangian dynamic makes the system to evolve on the constraint towards the minimum.
The best trade off is to choose a middle value (c=1). We approach the surface following the shape of the constraint. The number $\frac{1}{c}$ can be seen as the radius at which the lagrangian dynamics starts working. 
}{12}
\Example{}{Minimum distance from a point to a sphere
Consider the minimization problem, distance from a point to a sphere,
$$
min\ \frac{1}{2}(x^2+y^2+(z-2)^2) \ \ \ s.t. \ \ h(x)=x^2+y^2+z^2-1=0
$$
The solution is $(0,0,1)$. The projected system adding the sliding term is:
$$
\dot x = \nabla U +\lambda \nabla h=
\left( \begin{array}{ccccc} 
x+\lambda x+c(x^2+y^2+z^2-1)x \\
y+\lambda y +c(x^2+y^2+z^2-1)y\\
z-2+\lambda z +c(x^2+y^2+z^2-1)z\\
 \end{array} \right)
$$
with $\lambda=\frac{2z-1}{x^2+y^2+z^2}$. 
\begin{figure} [ht]
\centering
\includegraphics[scale=0.75]{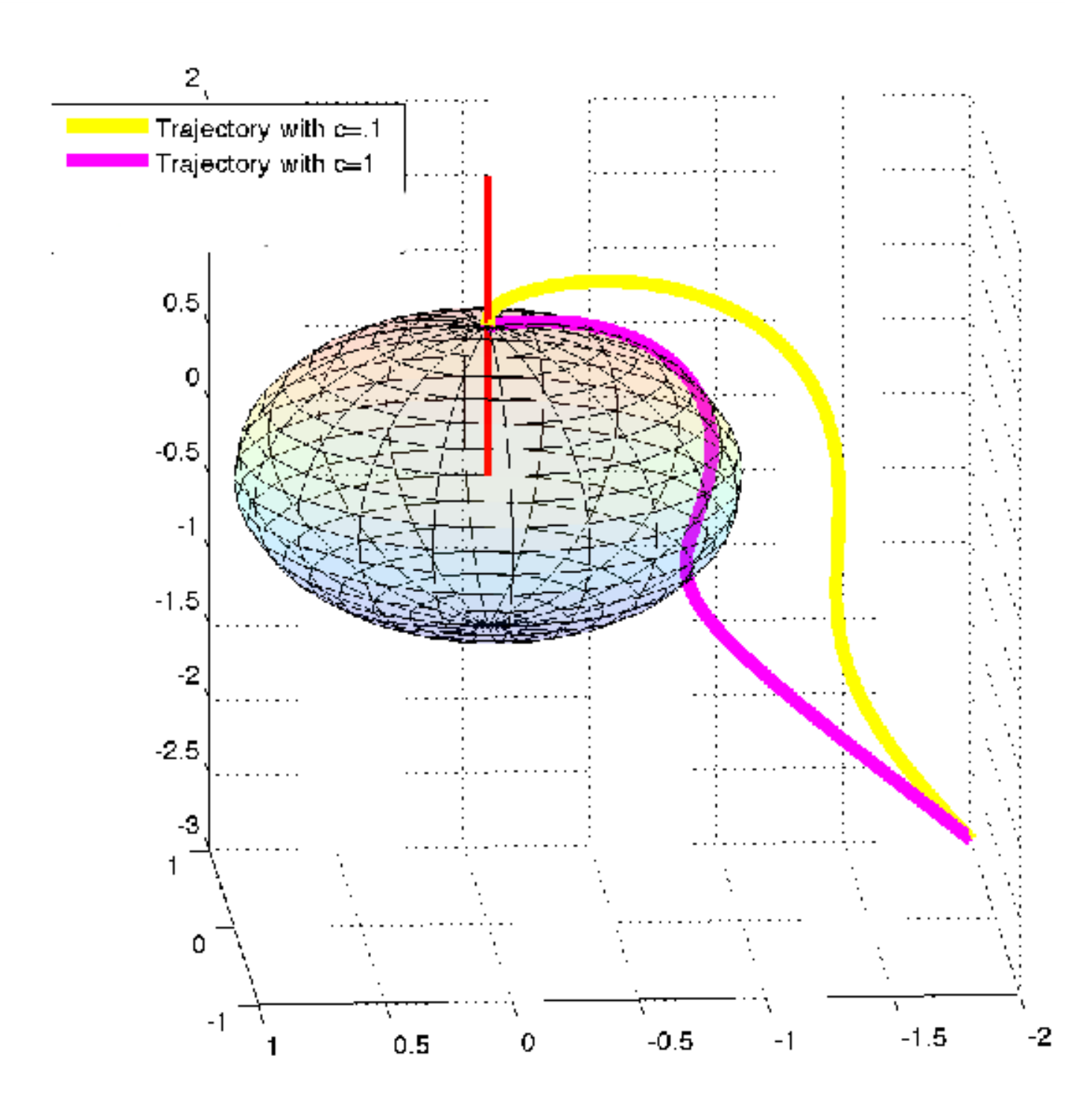}
\caption{Minimum distance from a point to a sphere}
\label{fig:sphere1sliding}
\end{figure}


}{13}

\Example{}{The problem is to maximize the volume of a parrallelepiped staying on an ellipse
$$
min -xyz \ \ \ s.t. \ \ h(x)=\frac{x^2}{a^2}+\frac{y^2}{b^2}+\frac{z^2}{c^2}-1=0
$$
Consider $a=15;b=5;c=3;$. figure \ref{fig:ellipseaugmented}  shows the convergence. 
\begin{figure} [ht]
\centering
\includegraphics[scale=0.75]{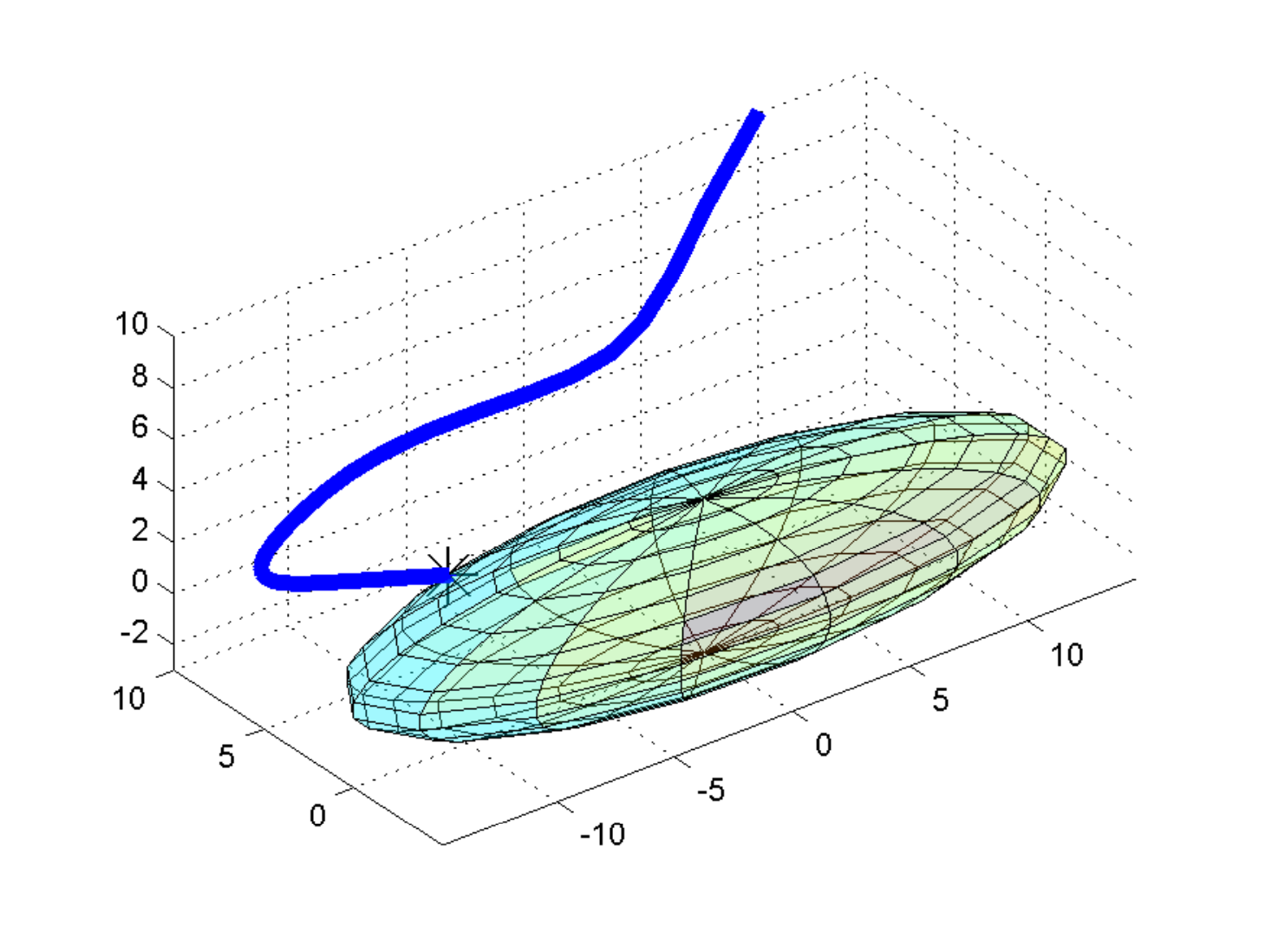}
\caption{Maximum volume on an ellipse}
\label{fig:ellipseaugmented}
\end{figure}

}{14}
\Example{}{
The minimization problem with two constraints
$$
min\ \frac{1}{2}||{\bf x}-{\bf x_1}||^2  \ \ \ s.t. \ \ h_1({\bf x})=||{\bf x}||^2-1=0 \ \ h_2({\bf x})=x+y+z=0
$$
The final point is the double projection of the absolute minimum. The values $c_1, c_2$ are the gains to the two surfaces. To stay on the sphere, we increase the gain ($c_1=10$).
\begin{figure} [ht]
\centering
\includegraphics[scale=0.75]{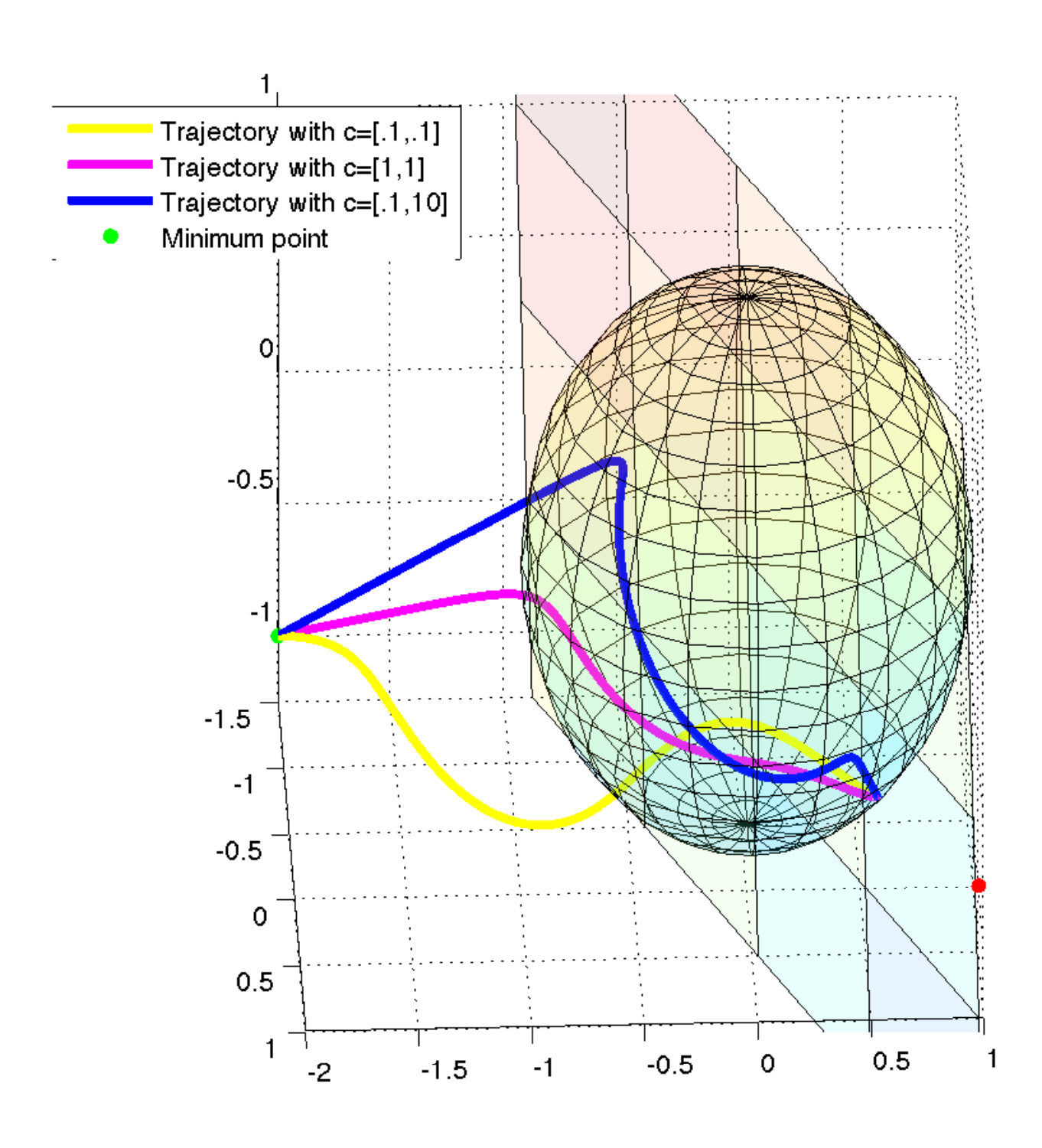}
\caption{Example of minimization with two constraints}
\label{fig:twoconstraintssliding}
\end{figure}
}{15}
\section{Contraction theory with time-varying equality constraints}
This section has two main parts. In the first one, we state a theorem giving the condition of contraction for a time varying constrained nonlinear dynamic system. We start on the constraints to have contraction behavior. In the second part, we allow the system to start outside the constraints. We use sliding techniques to conclude convergence to the constraints.
\subsection{Starting on the constraints: contraction theory}
We define the set $S({\bf x},t) = [ {\bf x}\ / \ {\bf h}({\bf x},t)=0]$.
\subsubsection{Contraction theorem}
\begin{figure} [ht]
\centering
\includegraphics[scale=0.75]{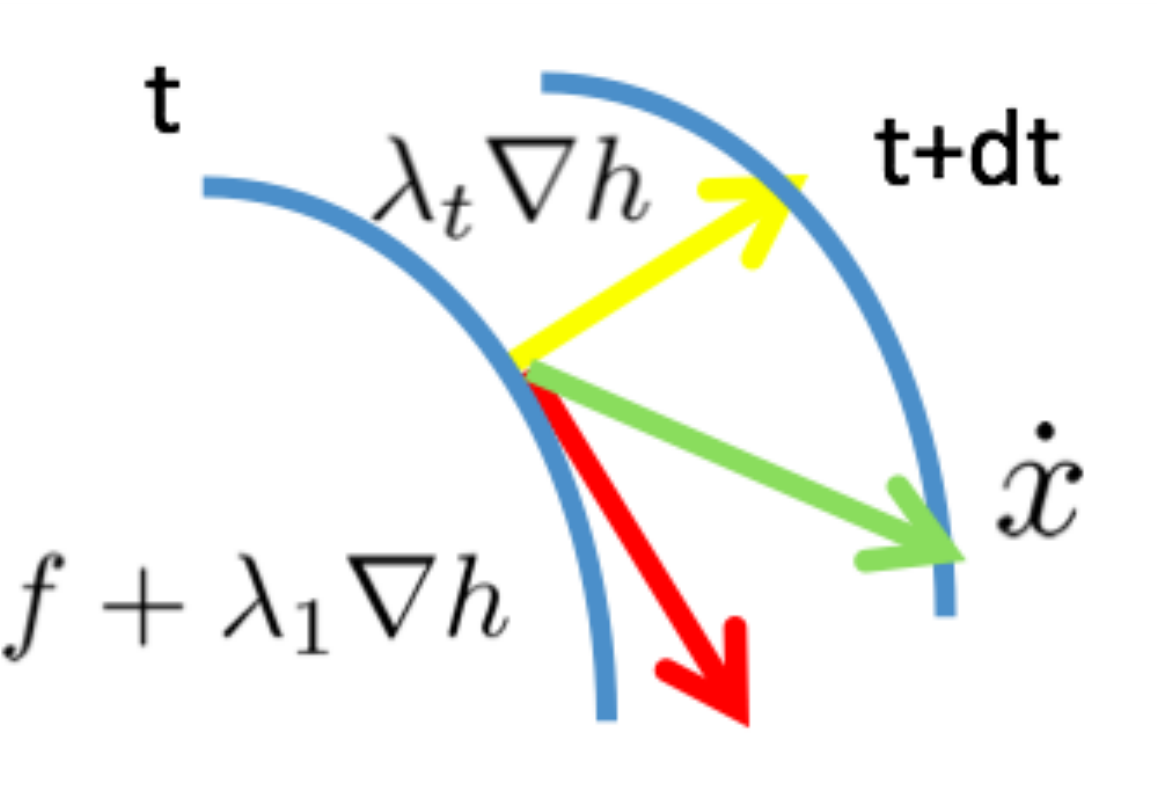}
\caption{Projection with time-varying constraints}
\label{fig:time}
\end{figure}

In this section we compute the condition of contraction for dynamic systems under time-varying constraints.
\begin{theorem}\label{timevarying}
The condition of contraction for the system $$\dot {\bf x} = {\bf f}({\bf x},t)  \  s.t. \ {\bf h}({\bf x},t)=0$$ is that there exists a uniform positive definite metric; $\Theta$ and that the hermitian part of :
$$
{\bf F} = \left (\dot {\bf \Theta}({\bf x},t)  +  {\bf \Theta}({\bf x},t)P({\bf x})\left (\frac{\partial{\bf f}}{\partial{\bf x}}+\lambda({\bf x},t)\frac{\partial ^2{\bf h}({\bf x})}{\partial{\bf x}^2}\right ) \right ) {\bf \Theta}({\bf x},t)^{-1} \ \ on \ S({\bf x},t)
$$
has a maximum eigenvalue uniformly negative. $P({\bf x})$ is the projection operator onto the tangent space $S({\bf x},t)$ and $\lambda({\bf x},t)=-({{\bf f}({\bf x},t)\nabla {\bf h}+\frac{\partial {\bf h}}{\partial t})[}{\nabla {\bf h}\nabla {\bf h}'}]^{-1}$. The initial condition must verify the constraint.
\end{theorem}
The proof has three parts. The initial problem is 
$$\dot {\bf x} = {\bf f}({\bf x},t)  \  s.t.  \ h({\bf x,t})=0$$
Physically to remain on the constraints, we need to project the speed and consider the surface's speed, figure \ref{fig:time}
This last term is calculated because the constraints are always verified,
$
0=\frac{dh}{dt}=\nabla h \dot x' + \frac{\partial h}{\partial t}
$
As we have the choice on $\dot x'$, we add a term to ensure that this equation is verified.
$$
-\frac{\partial {\bf h}}{\partial t}[\nabla {\bf h} \nabla {\bf h}']^{-1}\nabla {\bf h} = \lambda_t \nabla {\bf h}
$$
We can see this new parameter as a force that makes the system stay on the constraint. It is a new generalization of Lagrange parameters.
The total system is:
$$
\dot x = {\bf f}({\bf x},t) + \lambda_1 \nabla {\bf h}  -\frac{\partial {\bf h}}{\partial t}[\nabla {\bf h} \nabla {\bf h}']^{-1}\nabla {\bf h} = {\bf f}({\bf x},t) + \lambda_1 \nabla {\bf h} + \lambda_t \nabla {\bf h} = f(x,t)+\lambda \nabla {\bf h}
$$
where $\lambda=\lambda_1+\lambda_t$. The constraints are verified at anytime because the system starts on the constraint and the velocity remains on the constraints. 

We compute $ \delta \dot{{\bf x} } =\left ( \frac{\partial f}{\partial{\bf x}}+\lambda({\bf x})\frac{\partial ^2h}{\partial{\bf x}^2}+ \frac{\partial \lambda_1({\bf x})}{\partial{\bf x}}\frac{\partial h}{\partial{\bf x}} + \frac{\partial\lambda_t({\bf x})}{\partial{\bf x}}\frac{\partial h}{\partial{\bf x}}\right ) \delta {\bf x}$. $\frac{\partial\lambda_1({\bf x})}{\partial d{\bf x}}$ can be computed
as well as $\frac{\partial \lambda_t({\bf x})}{\partial{\bf x}}$.

We can substitute in the precedent equation and reorganize:
$$
\delta \dot{{\bf x} } =  P({\bf x})\left (\frac{\partial f}{\partial {\bf x}}+\lambda({\bf x})\frac{\partial ^2h}{\partial {\bf x}^2}\right ) \delta {\bf x} 
-{\nabla h({\bf x})'}[{\nabla h({\bf x}) \nabla h({\bf x})'} ]^{-1} \left (\delta {\bf x}\frac{\partial ^2h}{\partial {\bf x}^2}\dot {\bf x}+\frac{\partial h}{ \partial t}\right )
$$
It is essential to note that the first term belongs to $M({\bf x})$ and the second term belongs to $M({\bf x})^{\perp}$. To have contraction behavior the first term of $\delta \dot{{\bf x}^{\parallel} }$ must be uniformly bounded. 
$$
\delta \dot{{\bf x} } =  P({\bf x})\left (\frac{ \partial f}{\partial {\bf x}}+\lambda({\bf x})\frac{ \partial ^2{\bf h}}{\partial {\bf x}^2}\right ) \delta {\bf x} 
$$
Finally 
$$
\frac{d}{dt}(\delta {\bf z}' \delta {\bf z} ) = 2\delta {\bf z}' \left(\dot \Theta({\bf x},t) +  \Theta({\bf x},t)  P({\bf x}) \left(\frac{\partial f}{\partial{\bf x}}+\lambda({\bf x})\frac{\partial ^2{\bf h}}{\partial{\bf x}^2}\right ) \right ) \Theta({\bf x},t)^{-1}\delta {\bf z} 
$$
\subsubsection{Theorem's applications}
One of the most important application is optimization under varying constraints.
\Example{}{
The problem of the sum of the sides under a growing circle is :
$$
min \ x+y \ s.t. \ x^2+y^2=t^2
$$
The solution using KKT theorem, \ref{th:constrained}, is given by $ x^*=y^*=-\frac{t}{\sqrt{2}}$. Using the precedent theorem we create the 'extended' lagrangian system, simulated on figure \ref{fig:varyingconstraint}
$$\dot x=-
\left(
\begin{array}{cc}
  1-\frac{(x+y)}{t^2}x-\frac{x}{t}\\
  1-\frac{(x+y)}{t^2}y-\frac{y}{t}\\
\end{array}
\right)
$$
The condition of contraction is $\nabla^2f+(\lambda_1+\lambda_t)\nabla^2h = -\frac{x+y+t}{t^2}I > 0 $. The final condition is 
$x+y+t<0$. 
\begin{figure} [ht]
\centering
\includegraphics[scale=0.75]{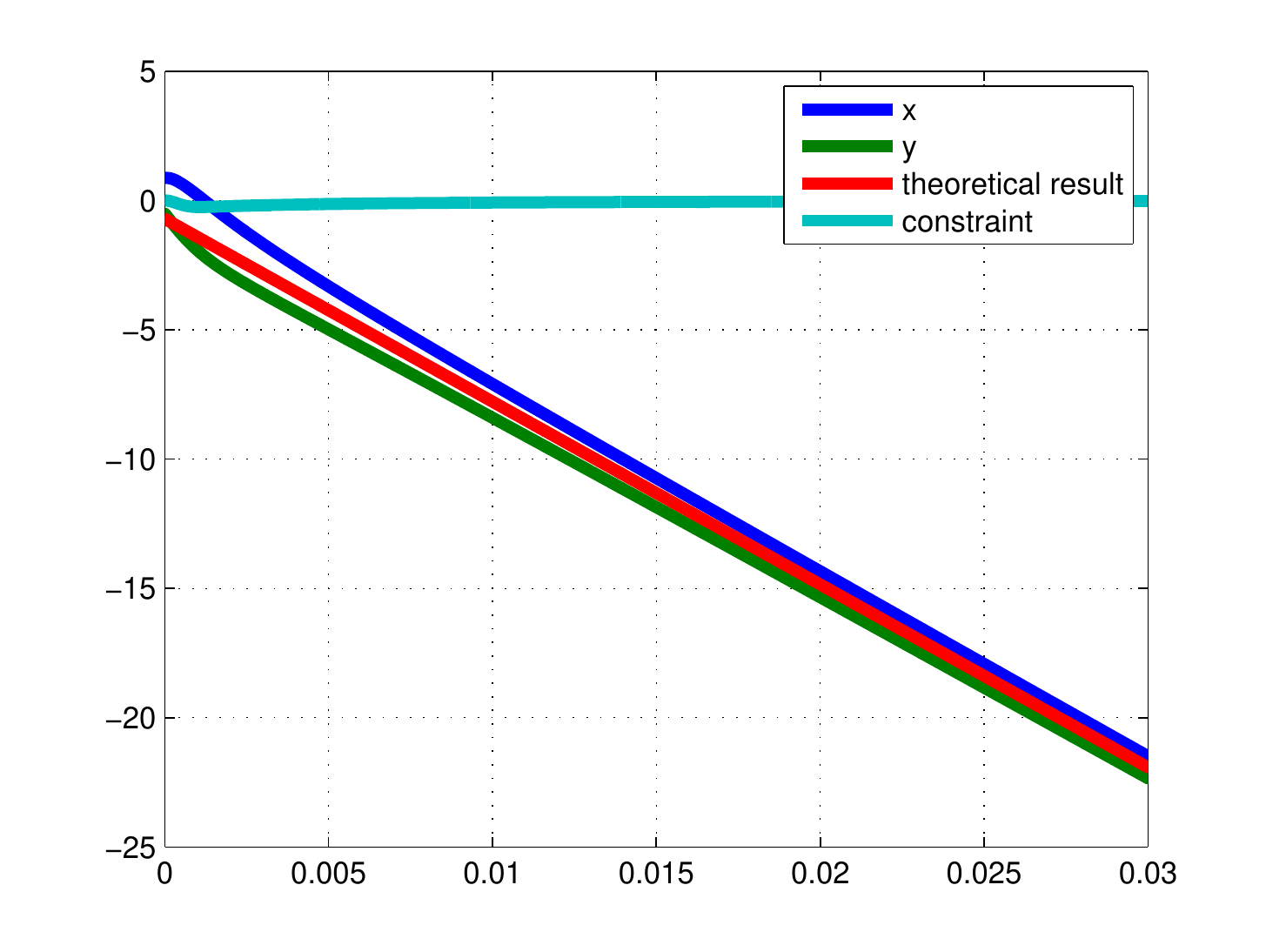}
\caption{Minimization of the sum of the sides under a growing circle}
\label{fig:varyingconstraint}
\end{figure}
}{16}

\Example{}{
The problem of the sum of the sides under a growing ellipse is :
$$
min \ x+y \ s.t. \ tx^2+y^2=1
$$
The solution using KKT theorem, \ref{th:constrained}, is given by $ x^*=-\frac{1}{\sqrt{1+t^2}}$, $y^*=-\frac{t}{\sqrt{1+t^2}}$.
$$\dot x=-\left(
\begin{array}{cc}
  1-\frac{(tx+y)}{t^2x^2+y^2}tx-\frac{x^2}{2t^2x^2+2y^2}xt\\
  1-\frac{(tx+y)}{t^2x^2+y^2}y-\frac{x^2}{2t^2x^2+2y^2}y\end{array}
\right)
$$
The condition of contraction is given by $\nabla^2f+(\lambda_1+\lambda_t)\nabla^2h = -\frac{x^2+tx+y}{t^2x^2+y^2}I > 0 $. The final condition is $x^2+tx+y<0$. 
\begin{figure} [ht]
\centering
\includegraphics[scale=0.75]{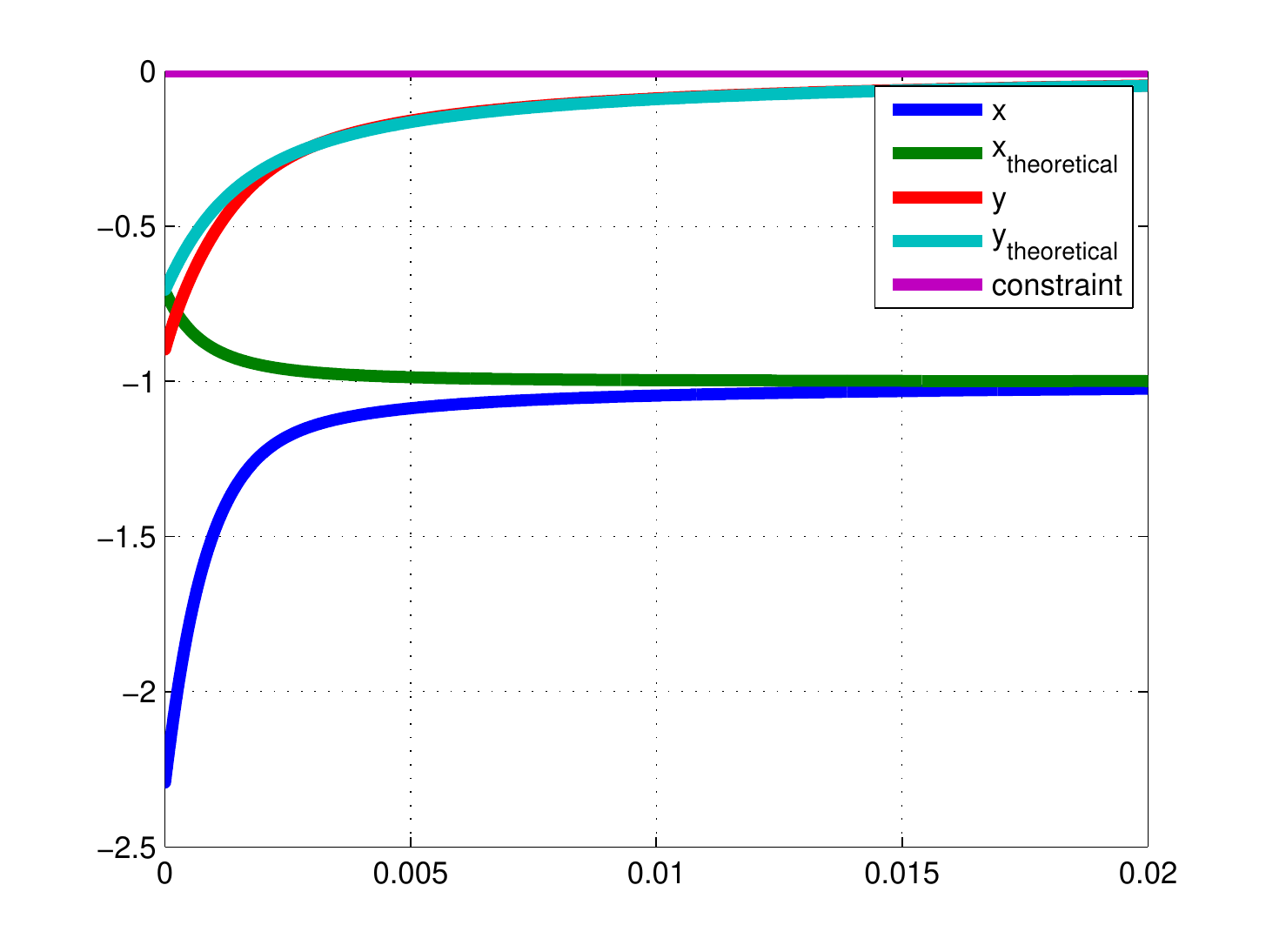}
\caption{Minimization of the sum of the sides under a growing ellipse}
\label{fig:varyingconstraints2}
\end{figure}
}{17}
In this example; we use time varying constraints to solve problems with static constraints.
\Example{}{
In order to make a smoother approach to a certain constraint $h_0$, we can create a time varying constraint such that when $t \rightarrow \infty$, $h(x,t) \rightarrow h_0(x). $ Consider the minimization problem:
$$ min\ x+y \ \ s.t. \ x^2+y^2=e^{\frac{2}{t}}$$
The dynamical system associated to that is :

$$\dot x=-\left(
\begin{array}{cc}
  1-\frac{(x+y)}{e^{\frac{2}{t}}}x-\frac{x}{t^2}\\
  1-\frac{(x+y)}{e^{\frac{2}{t}}}y-\frac{y}{t^2}\\\end{array}
\right)
$$
\begin{figure} [ht]
\centering
\includegraphics[scale=0.75]{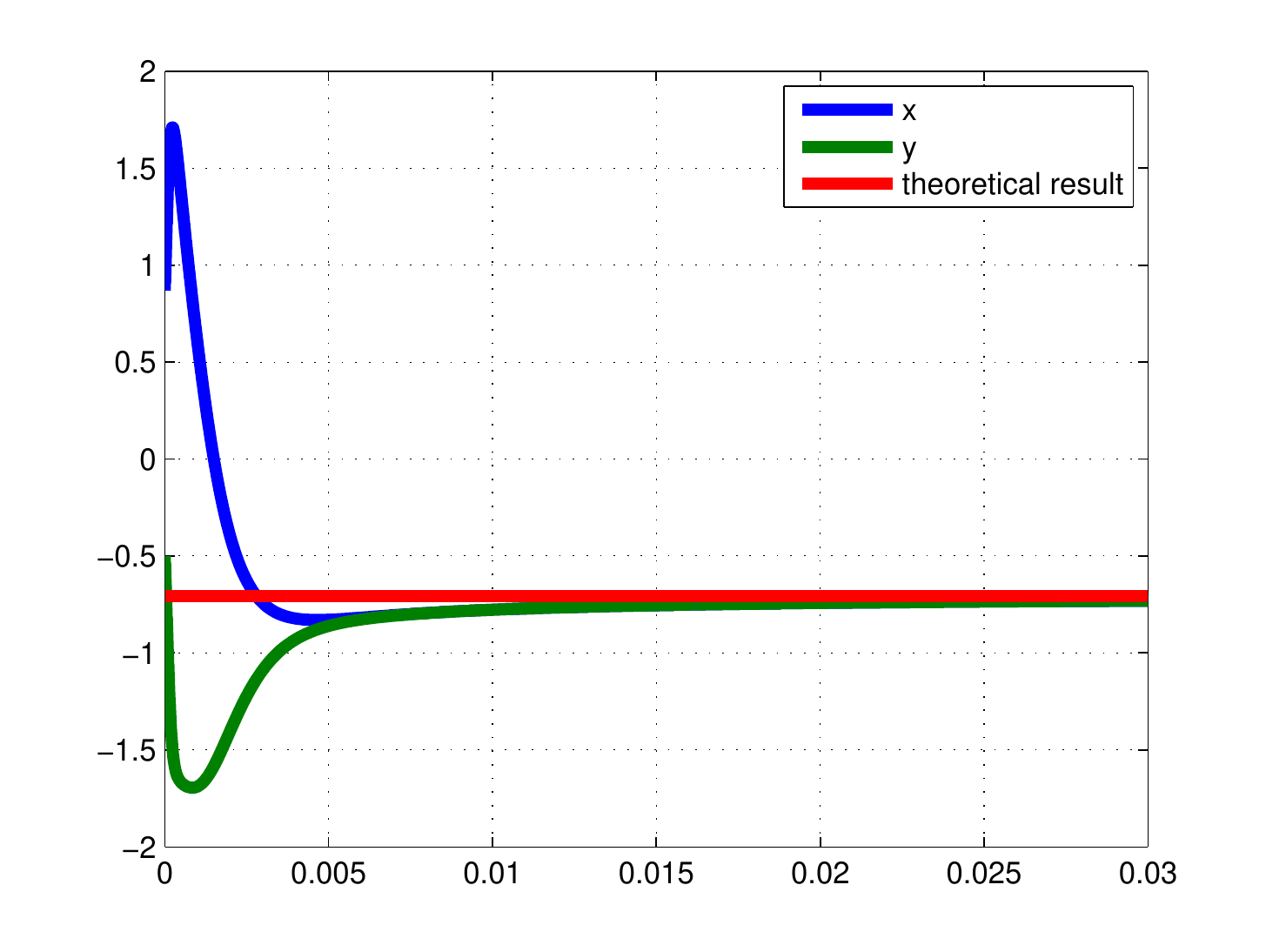}
\caption{Using time varying constraints to solve problems with static constraints}
\label{fig:varyingconstraints3}
\end{figure}
}{18}
\subsection{Starting outside the constraints: sliding behavior}
\subsubsection{Sliding surface behavior}
As done in the precedent chapter, we add a sliding term. The new system is :
$$
\dot x =  {\bf f}(x,t)+\lambda \nabla {\bf h} -\frac{\partial {\bf h}}{\partial t}[\nabla {\bf h} \nabla {\bf h}']^{-1}\nabla {\bf h}- \sum_j c_j h_j(x)\nabla h_j(x) 
$$ 
We have global convergence towards the surface. We define $s=\frac{1}{2}(\sum_i c_i h^2_i(x))$ where $c_i>0$. ${\bf s}$ is by definition strictly positive. Its derivative:
$$
\frac{d{\bf s}}{dt}=C'{\bf h}' \left( \frac{\partial {\bf h}}{\partial x}\dot x' + \frac{\partial {\bf h}}{\partial t}\right )
 $$
 $$
 = C'{\bf h}' \left( \frac{\partial {\bf h}}{\partial x}    \left( P(x)'{\bf f}(x,t)'-\nabla {\bf h}'[\nabla {\bf h} \nabla {\bf h}']^{-1}\frac{\partial {\bf h}}{\partial t}- \sum_j c_j h_j(x)\nabla h_j(x)  \right)  + \frac{\partial {\bf h}}{\partial t}\right ) 
$$
Because ${\bf f}(x,t)P(x) \in M(x)$, $\nabla h_i(x){\bf f}(x,t)P(x) = 0$. As $\nabla {\bf h} \nabla {\bf h}'[\nabla {\bf h} \nabla {\bf h}']^{-1}\frac{\partial {\bf h}}{\partial t}=\frac{\partial {\bf h}}{\partial t}$. The derivative is
$$
\frac{ds}{dt}=-(\sum c_i h_i \nabla h_i)^2 \le 0
$$
The second derivative is smooth, it is bounded. 
Applying Barbalat's lemma, we conclude that $\sum c_i h_i \nabla h_i$ converges to zero as time goes to infinity. As $\nabla h_i $ are linearly independent, $c_i h_i$ goes to zero. The constraints are verified.

\subsubsection{Examples}
To show how this sliding term works, we use the precedent examples but starting on a point outside the surface.
\Example{}{Minimization of the sum of the sides under a growing circle
The problem is :
$$
min \ x+y \ s.t. \ x^2+y^2=t^2
$$
The new system is
$$\dot x=-
\left(
\begin{array}{cc}
  1-\frac{(x+y)}{x^2+y^2}x-\frac{x}{t}+chx\\
  1-\frac{(x+y)}{x^2+y^2}y-\frac{y}{t}+chy\\
\end{array}
\right)
$$
\begin{figure} [ht]
\centering
\includegraphics[scale=0.75]{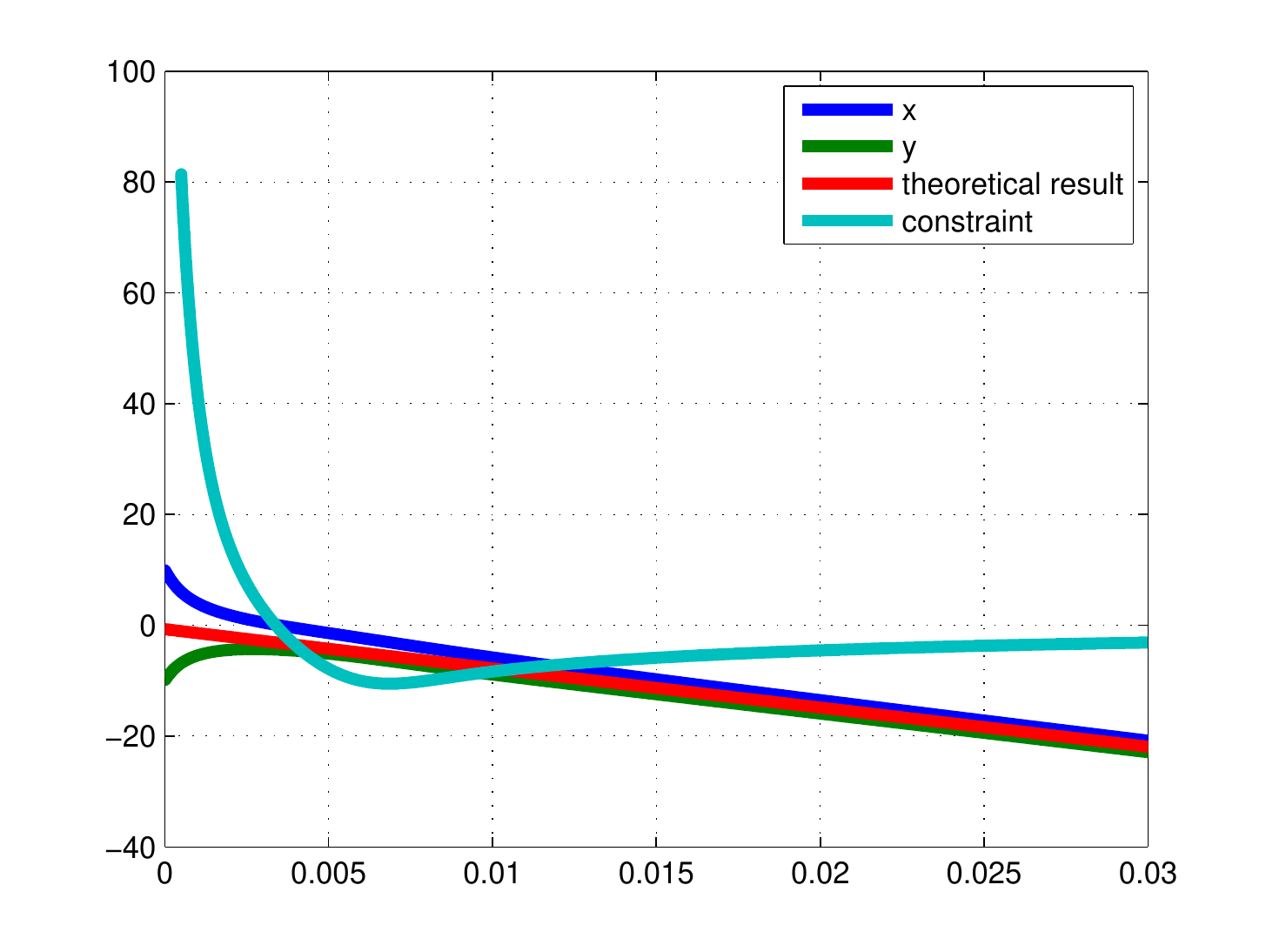}
\caption{Minimization of the sum of the sides under a growing circle}
\label{fig:varyingconstraints4}
\end{figure}
}{19}
\Example{}{Minimization of the sum of the sides under a growing ellipse
The problem is :
$$
min \ x+y \ s.t. \ tx^2+y^2=1
$$
The new system is
$$
\dot x=-\left(
\begin{array}{cc}
  1-\frac{(tx+y)}{t^2x^2+y^2}tx-\frac{x^2}{2t^2x^2+2y^2}xt+chxt\\
  1-\frac{(tx+y)}{t^2x^2+y^2}y-\frac{x^2}{2t^2x^2+2y^2}y+chy
\end{array}
\right)
$$
\begin{figure} [ht]
\centering
\includegraphics[scale=0.75]{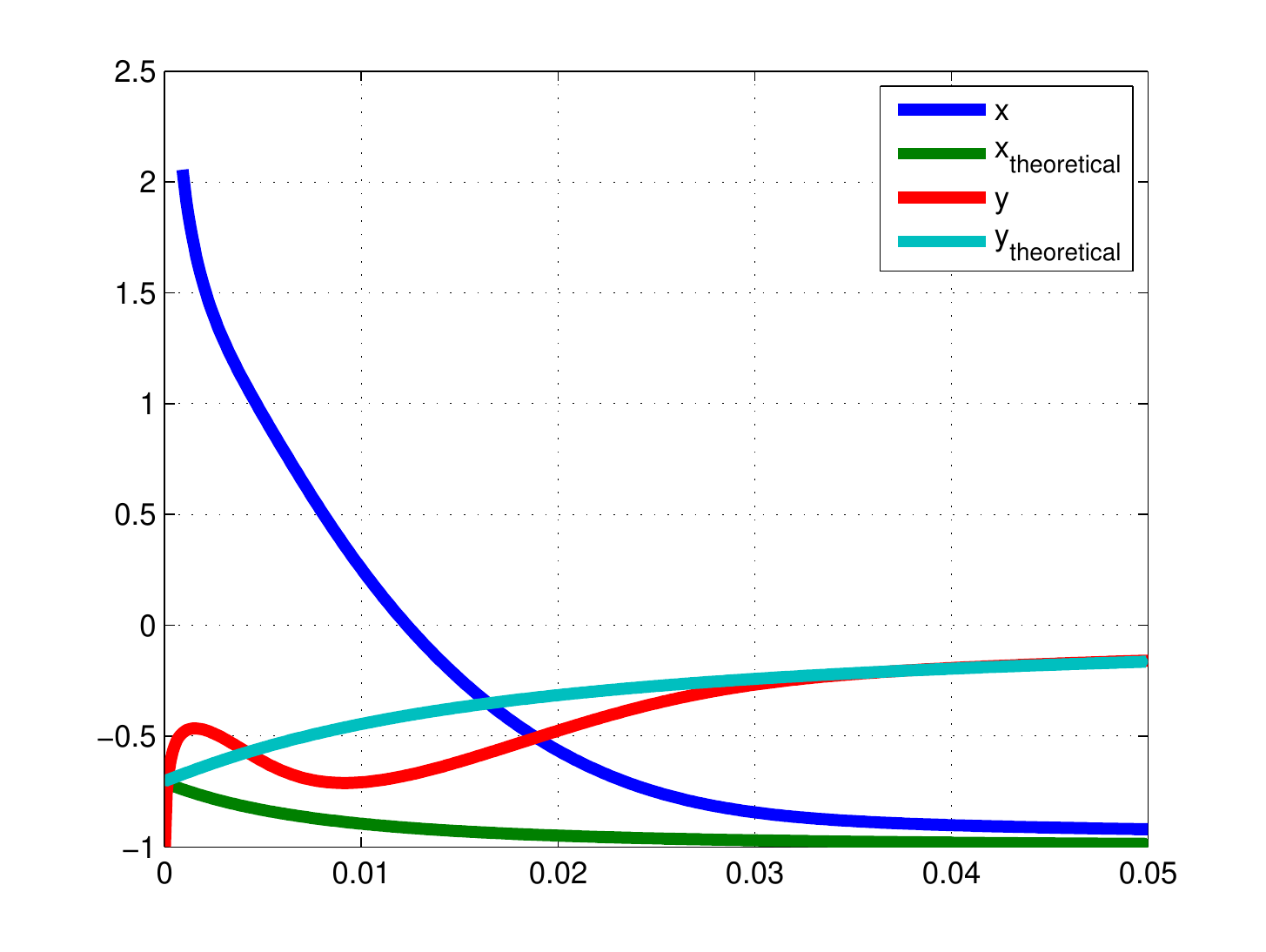}
\caption{Minimization of the sum of the sides under a growing ellipse}
\label{fig:varyingconstraints5}
\end{figure}

}{20}

\section{Conclusion and directions for future research}
This research has four main new contributions. 

\begin{itemize}
\item[-] The first and more important is the condition of contraction (a condition of convergence) for two different dynamic systems, constrained with equalities and constrained with time varying inequalities.
\item[-]  The second contribution is, in the particular case of a gradient autonomous contracting in a metric identity system, an algorithm to find the minimum. A more general metric could allow solving more complicated problems.
\item[-]  The third contribution is the understanding that adding a term $\frac{h^2}{2}$ in the cost function corresponds to dynamic term that makes the dynamic system converge towards the surface. 
\item[-]  The fourth and last contribution is on Lagrange parameters. This paper gives a new physical approach of Lagrange parameters. They can be understood as a scalar value that ensure the system to have a tangential speed to the constraints. Also it gives three generalizations of Lagrange parameters, with a time varying cost function $\lambda=-\nabla U({\bf x},t)\nabla h'[\nabla h\nabla h']^{-1}$, with a time varying general vector $\lambda=-{\bf f}({\bf x},t)\nabla h'[\nabla h\nabla h']^{-1}$ and with a a time varying general vector with time varying constraints $\lambda=-\left({\bf f}({\bf x},t)\nabla h'+\frac{\partial h}{\partial {\bf x}}\right)[\nabla h\nabla h']^{-1}$. This last case is the most general because it contains the other two.
\end{itemize}

There are many directions for the future research. 
\begin{itemize}
\item[-]  We have always used an identity metric. It is interesting to explore what happens with constant metrics and time varying metrics. In the gradient autonomous case, it could include some information of the constraints making the minimization faster.
\item[-]  When having many different goals, Pareto optimality does not have a general theorem to find the best solution. May be some conclusive theorem can be found exploring Pareto optimality using contraction theory. Contraction theory has very interesting combination properties. 
\item[-]  In chapter one, an example using parallel combination has been presented. We could try to use feedback and hierarchical combination. 
\item[-]  The Hamilton-Jacobi-Bellman (HJB) equation is a partial differential equation which is central to optimal control theory. It is the solution of a minimization problem subject to a dynamic system. 
\item[-]  In the article $\cite{amari}$, the ordinary gradient of a function does not represent its steepest direction, but the natural gradient does. Information geometry is used for calculating the natural gradients. How does this natural gradient relates to our projection operator ?.
\end{itemize}

\bibliographystyle{IEEEtran}


\end{document}